\documentclass{aastex631}

\newcommand\etain{\eta_{\text{in}}}
\graphicspath{{./}{./figures/}}
\usepackage{physics}
\usepackage{derivative}
\usepackage{mathrsfs}
\usepackage{placeins}
\usepackage{subfigure}

\usepackage{xcolor}

\newcommand{\vv}{\mathbf{v}}
\newcommand{\vU}{\mathcal{U}}
\newcommand{\vin}{v_\text{in}}
\newcommand{\rin}{r_\text{in}}
\newcommand{\rhoin}{\rho_\text{in}}
\newcommand{\Mdotin}{{\dot{M}}_\text{in}}
\newcommand{\Rs}{R_\text{s}}
\newcommand{\vs}{v_\text{s}}
\newcommand{\Rsw}{R_\text{sw}}
\newcommand{\vsw}{v_\text{sw}}
\newcommand{\Rc}{R_\text{c}}

\newcommand{\rhoa}{\rho_\text{a}}
\newcommand{\rhow}{\rho_\text{w}}
\newcommand{\lfrac}[2]{{#1}/{#2}}
\newcommand{\vratio}{\lfrac{\vin}{\vsw}}
\newcommand{\awind}{a_\text{wind}}
\newcommand{\aamb}{a_\text{amb}}
\newcommand{\xsw}{x_\text{sw}}
\newcommand{\xs}{x_\text{s}}
\newcommand{\uin}{u_\text{in}}
\newcommand{\gammasa}{\gamma_\text{sa}}
\newcommand{\gammasw}{\gamma_\text{sw}}

\newcommand{\cm}{\,\text{cm}}
\newcommand{\second}{\,\text{s}}
\newcommand{\km}{\,\text{km}}
\newcommand{\cms}{\cm\second^{-1}}
\newcommand{\kms}{\km\second^{-1}}
\newcommand{\gram}{{\mathrm{\,g}}}
\newcommand{\massden}{{\gram\cm^{-3}}}
\newcommand{\rsqmassden}{{\gram\cm^{-1}}}
\newcommand{\yr}{\,\text{yr}}
\newcommand{\au}{\,\text{au}}

\begin{document}

\title{Polytropic Wind-Driven Bubbles and their Shock Structures in Radially Stratified Ambient Media}

\author[0009-0003-4528-1635]{Dmitrii Zagorulia}
\affiliation{Institute of Astronomy and Astrophysics, Academia Sinica, Taipei 106319, Taiwan}

\author[0000-0001-8385-9838]{Hsien Shang}
\affiliation{Institute of Astronomy and Astrophysics, Academia Sinica, Taipei  106319, Taiwan}
\correspondingauthor{Hsien Shang}
\email{shang@asiaa.sinica.edu.tw}

\author[0000-0001-5557-5387]{Ruben Krasnopolsky}
\affiliation{Institute of Astronomy and Astrophysics, Academia Sinica, Taipei  106319, Taiwan}

\begin{abstract}
We extend the analytic expressions for polytropic wind-driven bubbles and their shock structures, formulated initially in Koo and McKee 1992(a,b), focusing on spherically symmetric configurations in astrophysical environments with $\rho\propto r^{-2}$, which arises naturally in the star-forming environment and has applications to winds flowing into a preexisting bubble.
Wind luminosity is assumed to be constant, and as a result the shock velocities of these bubbles are constant in time. The ratio of specific heats is assumed to be the same in the shocked ambient medium and the shocked wind. Numerical results are presented for one selected ratio of wind density to ambient density.
Exact ODEs are written for the compressed wind region and approximate solutions are found by fitting the ODE solutions. By analyzing the interactions between stellar winds and ambient media in the strong compression limit, we model the formation and evolution of spherical bubbles, highlighting their shock fronts and contact discontinuities. Our analytic method provides an intuitive approach to calculating the thickness of bubble shells, which is crucial for understanding their dynamics and observational characteristics.
A numerical method explores conditions without explicitly requiring the strong compression limit, and then we compare numerical to analytical results under various conditions.
\end{abstract}


\section{Introduction} \label{sec:intro}
For several decades, wind-blown bubbles around massive stars have been the subject of extensive research in astrophysics. These structures, formed by the interaction between stellar winds and the surrounding interstellar medium (ISM), play a crucial role in shaping the dynamics and evolution of star-forming regions. Moreover, the bubble structures are not limited to winds from massive stars or star-forming environments. Examples of the interactions between powerful winds and an environment exist for supernovae, pulsar winds, and galactic-scale processes.

The detailed study of wind-blown bubbles traces its roots to the pioneering work of \citet{weaver1977}, who established the fundamental analytical framework for understanding these structures. Subsequent contributions, including the seminal works of \citet{koo_mckee_i,koo_mckee_ii}, have significantly expanded our knowledge of bubble dynamics and their interaction with the ambient medium. This theory builds upon the previous work on blast waves \citep{taylor_i,taylor_ii,sedov1946,zeldovich1967}, which can be seen as a single-pulse thermally-driven analog of the cold and usually more time-extended wind-driven bubbles. 
More recently, \citet{lancaster2021_i,lancaster2021_ii} have explored the effects of turbulence on bubbles, finding that cooling can be enhanced due to the fractal complexity of the surface in turbulent conditions.
The theory of Protoplanetary Nebulae (PPN) and Planetary Nebulae (PN) as recently developed, e.g., by \citet{garcia-segura2018,garcia-segura2020,garcia-segura2021,garcia-segura2022} takes concepts and applications from wind-blown bubble theory, with similar basic shock structures of wind and ambient shocks.
The common envelope of the binary stars evolves \citep{ricker2012} to create an envelope, 
and an accretion disk able to produce wide-angle jets and winds through magneto-centrifugal mechanisms \citep{blandford1982,shu1994_XWI}, such as collimated X-winds \citep{shu1995}. These winds blow bubbles through the material of the ejected envelope and beyond, forming the basic structure of the PN\@.

Wind-blown bubbles are expanded beyond the galactic scale in the form of superbubbles and inter-scale feedbacks, where the collective effects of multiple massive stars in OB associations create larger structures that play a crucial role in galactic feedback processes and the enrichment of the intergalactic medium. Multi-wavelength observations have allowed for detailed studies of the hot gas and dust within bubbles. 

Wind-driven bubbles typically exhibit a characteristic structure consisting of several distinct regions. Going from the inside to the outside, those are four regions, separated by three discontinuities:
(1) free-flowing wind region,
(2) reverse shock (wind termination shock),
(3) hot region of shocked wind,
(4) contact discontinuity,
(5) hot region of shocked ambient material,
(6) forward shock, and, surrounding the whole bubble structures, 
(7) largely unperturbed ambient material.
The evolution of these bubbles is governed by complex interactions between the winds and the ambient media, leading to a rich set of physical phenomena through various instabilities and mixing structures. High-resolution simulations have shed light on various instabilities that can develop in wind-blown bubbles, including Rayleigh--Taylor, Kelvin--Helmholtz, and thermal instabilities. These processes significantly mix different bubble regions, affecting their structure and evolution.

The classical outflow structures blown by winds from young stellar objects (YSOs) have been recognized as a special bubble structure through magnetohydrodynamic (MHD) simulations, revealing complex field configurations and their impact on bubble morphology and expansion rates. While very elongated and strongly magnetized, they present the same basic structures of the interaction of wind with its environment as the more spherical bubbles, in addition to unique features of instabilities arising from the magnetized interplay of wind with its environment \citep{shang2020,shang2023_signatures}, extended beyond the very elongated hydrodynamic model in the classical momentum-conserving outflow thin shell \citep{shu1991,shang2023_signatures}. These bubbles' elongated and oblique shape raises the possibility of Kelvin--Helmholtz instabilities. Synthetic observations based on analytical models and simulations \citep{shang2023_signatures} serve to understand features of observations of YSO outflows as bubble structures of nested shells
\citep{shang2023_signatures, ai2024,liu2025}, including direct and indirect effects of instabilities, detectable in position--velocity diagrams.

In this work, we build upon the original framework of spherical wind-blown bubbles as provided in \citet{koo_mckee_i,koo_mckee_ii} to add to the understanding of the shock structure of the polytropic wind bubbles.
We consider steady winds blowing into ambient media with $\rho \propto r^{-2}$. This choice has various astrophysical applications, for example bipolar outflows and fast winds blowing into the environment created by a preexisting slow wind.
The analytical and numerical solutions of this work use only one global polytropic $\gamma$, while the methods can be generalized to cover the cases in which $\gamma$ of the wind differs from that of the ambient shock.
These solutions with a single $\gamma$ do not apply to the common case of an adiabatic bubble with a radiative outer shock.

For the spherically symmetric cases, we developed a new method to obtain the relative thickness of regions filled up with shocked material from the wind and from the ambient medium, extending to the shocked wind region the analytical methods used for the shocked ambient medium in \citet{koo_mckee_ii} and \citet{weaver1977}, based on solving the basic ODE equations of the physics of the problem, and obtaining matching solutions that fulfill all the discontinuity conditions (Section \ref{sec:method}).
This new approach authentically solves the basic ODES of the shocked wind region, and it does not an the isobaric assumption for that calculation. In addition to that exact ODE approach, alternatively, we use a hyperbolic approximation for the ODE solutions of the compressed wind region, which yields the solution matching without the need of solving the hydrodynamic equations. The hyperbolic approximation is justified based on the shape of the profile of velocity as a function of position within the compressed wind region, as seen from the ODEs and from the numerics.
The analytical solution has been compared successfully with global 1D numerical simulations of the bubbles.
In Section \ref{sec:vin}, we explore with high-resolution 1D global simulations an expanded space of solutions and find that for small $\gamma$ (including intense cooling and the approach to the isothermal limit), the relative bubble thickness has a dependence on the wind velocity. This extends to the polytropic EOS some previous results obtained for the isothermal and two-temperature EOSs \citep{shang2020}.
In Section \ref{sec:vin_compare} we compare the results of the various methods. In Sections \ref{sec:applications_wind} and \ref{sec:applications_YSO} we discuss astrophysical applications. Conclusions are summarized in Section \ref{sec:conclusions}. Appendix \ref{sec:isothermal} covers a globally isothermal bubble, and Appendix \ref{sec:riemann} contains an analytical solution for an analog Cartesian Riemann problem.

\section{Methods and Results} \label{sec:all_results}

We now present the ODEs used to solve the shocked ambient medium (\S\ref{sec:shocked_ambient}) and the shocked wind region (\S\ref{sec:shocked_wind}) of the bubbles.
We describe the bubble structure with the usual hydrodynamic equations 
\begin{align}
    \pdv{\rho}{t} +  \nabla\cdot(\rho \vv) &= 0, \quad \text{(mass conservation)}\label{eq: hydro_i}\\
    \pdv{\vv}{t} + (\vv \cdot \nabla) \vv +\frac{1}{\rho} \nabla P &= 0,\quad \text{(momentum conservation)}\label{eq: hydro_ii}\\
    \pdv{E}{t} + \nabla\cdot(\vv(E + P)) &= 0, \quad \text{(energy conservation)}\label{eq: hydro_iii}
\end{align}
where $\rho$ is the density, $\vv$ is the velocity vector, and $P$ is the pressure. The total energy is $E = \rho \left(\vv^2/2 + e\right)$, where the specific internal energy $e$ is given by the equation of state (EOS).
We use this equation in spherical coordinates $(r, \theta, \phi)$.
The spherically symmetric bubble has all quantities depending only on $r$. We consider only the radial component of the velocity vector, which we call $v$.
Outside of discontinuities, the energy equation can be replaced by an entropy equation
\begin{equation}\label{eq:hydro_iv}
    \pdv{S}{t} + \vv\nabla S = 0\ .
\end{equation}
Here entropy can be (up to an additive constant) expressed as $S = C_V \ln(P \rho^{-\gamma})$, where $C_V$ is the specific heat at constant volume.

To set up the bubbles, we describe the initial wind and ambient medium as in KM92b. 
We consider in this work a wind with a constant velocity $\vin$,
\begin{equation}
    \Mdotin(t) = 4\pi \rhoin(t) \vin \rin^2\ ,
\end{equation}
where the mass density $\rhoin(t)$ is injected at an innermost boundary condition radius $\rin$, set to a value sufficiently small to not interfere with the analytical or numerical results. We consider in this work a steady wind and so fix $\etain = 1$, while in more general $\Mdotin \propto t^{\etain-1}$.
In Section \S\ref{sec:method}, we consider wind and ambient to be very cold so that the energy injection rate is fully kinetic and equal to $\mbox{\textonehalf}\Mdotin(t)\vin^2$, while in \S\ref{sec:vin} and Appendix \ref{sec:riemann} the wind is not necessarily completely cold.

We consider as initial condition a cold ambient medium at rest, with density ${\rhoa}(r) \propto r^{-k_\rho}$, with $k_\rho=2$.
The wind interaction with this ambient medium naturally generates a ``bubble.'' We have also  restricted our analysis to $\etain=1$ for the time dependence of the mass-injection rate,  which fulfills $\etain \le 3 - k_\rho$, corresponding to non-accelerated bubbles. The bubble structure has three discontinuities: an ambient shock at a radius $\Rs$, surrounding a contact discontinuity at $r=\Rc$, and a wind shock at a smaller radius $r=\Rsw$. 

At shocks, the flow must obey the Rankine-Hugoniot jump conditions \citep[e.g.,][]{shu1992}, which in the frame at rest with the shocks are the continuity of
mass, momentum, and energy
across the shock discontinuities.
At the contact discontinuity, density and temperature are discontinuous, but pressure and radial velocity are continuous.
The solution of the Rankine-Hugoniot jump conditions can be expressed in terms of the compression ratios for density and velocity
which takes the full form
\begin{equation}
\frac{\rho_\text{postshock}}{\rho_\text{preshock}}=\frac{u_\text{preshock}}{u_\text{postshock}}=
\frac{\gamma+1}{\gamma-1 + 2/M_\text{preshock}^2}\ , \label{eq:compression}
\end{equation}
where the velocity radial component $u$ is in the frame with moving at the shock speed. The preshock Mach number is $M_\text{preshock}=|u_\text{preshock}|/a_\text{preshock}>1$. In our case, at the ambient shock we have $M_\text{preshock}=|0-\vs|/\aamb$, and at the wind shock we have $M_\text{preshock}=|\vin-\vsw|/\awind$. The numerical results in section \ref{sec:vin} are based on a numerical code that approximates the full Rankine-Hugoniot conditions.
In section \ref{sec:method} we use the strong compression limit, in which the Rankine-Hugoniot compression ratio simplifies to
\begin{equation}
\frac{\rho_\text{postshock}}{\rho_\text{preshock}}=\frac{u_\text{preshock}}{u_\text{postshock}}=
\frac{\gamma+1}{\gamma-1}\label{eq:compression_limit}\ ,
\end{equation}
where the two preshock regions (free wind $r<\Rsw$, free ambient medium $r>\Rs$) obey respectively the conditions for wind and ambient media in their relevant cold limits.

\subsection{Semi-analytical Method} 
\label{sec:method}
The methods of this section \ref{sec:method} are based on KM92b. Section \ref{sec:shocked_ambient} reviews KM92b App.\ B.1, and section \ref{sec:shocked_wind} extends the methods to the shocked wind region.
Sections \ref{sec:matching} and \ref{sec:matched_solution} match the ambient and wind parts of the solution at the contact discontinuity, completing the semi-analytical method.

\subsubsection{Shocked Ambient Medium} \label{sec:shocked_ambient}
To make analytical progress on this problem, for the compressed ambient region $\Rc < r < \Rs$ treated in this section \ref{sec:shocked_ambient}, we review the results in Appendix B.1 of KM92b.
We assume that the self-similarity conditions have been achieved sufficiently well in late times. 
The front shock of the self-similar bubble expands in a medium with $\rhoa \propto r^{-k_\rho}$ as $\Rs \propto t^\eta$ \citep{ostriker1988}, where 
$\eta = \lfrac{(2 + \etain)}{(5 - k_\rho)}=\lfrac{(2 + 1)}{(5 - 2)}=1$.
The time-dependent position of the ambient shock can be used to define the dimensionless radial coordinate $\lambda \equiv \lfrac{r}{\Rs(t)}$ (a similarity variable).
Density, velocity, and pressure can be made dimensionless as new variables ($\Tilde{v}$, $\Tilde{P}$, $\Tilde{\rho}$), as defined in KM92b Eq.\ (B3):
\begin{equation}\label{eq: hydro_vars}
    \rho(r, t) = \rho_1(t) \Tilde{\rho}(\lambda), \quad v(r, t) = v_1(t) \Tilde{v}(\lambda), \quad P(r, t) = P_1(t) \Tilde{P}(\lambda)\ ,
\end{equation}
by using their postshock values at $\Rs$ as units.
The postshock values for density, velocity, and pressure could be expressed through the Rankine-Hugoniot jump conditions in the frame of reference moving at the shock velocity $\vs=d\Rs / dt = \eta \Rs / t$. Assuming $M \equiv M_\text{preshock}$,
\begin{equation}
    \rho_1(r) = \frac{\gammasa + 1}{\gammasa - 1 + 2/M^2} \rhoa(t), \quad v_1(t) = \frac{2 - 2 /M^2}{\gammasa + 1}\vs(t), \quad P_1(t) = \left(\frac{2 - 2/M^2}{\gammasa + 1} + \frac{1}{\gammasa M^2}\right)\rhoa(t) \vs^2(t).
\end{equation}
Let $\mu \equiv v_1/\vs$, one can obtain the system of ODEs for the shocked ambient medium, by substituting the dimensionless forms into 1D hydrodynamic equations in spherical coordinates (\ref{eq: hydro_i}-\ref{eq: hydro_iii}): 
\begin{align}
    \mu \dv{\Tilde{v}}{\lambda} + \left(\mu\Tilde{v} - \lambda\right) \frac{1}{\Tilde{\rho}} \dv{\Tilde{\rho}}{\lambda} + \frac{2 \mu \Tilde{v}}{\lambda} - k_\rho &= 0,\label{eq:ambient_1}\\
    \left(\mu\Tilde{v} - \lambda\right) \dv{\Tilde{v}}{\lambda} + \frac{P_1}{\rho_1 v_1 \vs} \frac{1}{\Tilde{\rho}} \dv{\Tilde{P}}{\lambda} + \left(1 - \frac{1}{\eta}\right)\Tilde{v} &= 0,\label{eq:ambient_2}\\
    \left(\mu\Tilde{v} - \lambda\right)\left(\frac{-\gammasa}{\Tilde{\rho}} \dv{\Tilde{\rho}}{\lambda} + \frac{1}{\Tilde{P}} \dv{\Tilde{P}}{\lambda} \right) + k_\rho (\gammasa -1) + 2 \left(1 - \frac{1}{\eta}\right) &= 0,\label{eq:ambient_3}
\end{align}

One can note that at $M \to \infty$ limit, $\mu \to 2/(\gammasa + 1)$, $P_1/(\rho_1 v_1 \vs) \to (\gammasa - 1)/(\gammasa + 1)$,
which exactly coincides with Eq.\ B5 of KM92b. Rewriting (\ref{eq:ambient_1}-\ref{eq:ambient_3}) explicitly for derivatives
\begin{align}
    \dv{\Tilde{v}}{\lambda} &= \left\{\frac{2 \gammasa \mu \Tilde{v}}{\lambda} - k_\rho - \left(1 - \frac{1}{\eta}\right) \left[C_\text{sa} \frac{\Tilde{\rho}\Tilde{v}}{\Tilde{P}}(\mu \Tilde{v} - \lambda) - 2 \right]\right\} \cdot \left[C_\text{sa} \frac{\Tilde{\rho}}{\Tilde{P}}(\mu \Tilde{v} - \lambda)^2 - \gammasa \mu\right]^{-1},\\
    \dv{\Tilde{P}}{\lambda} &= \left[-\left(1 - \frac{1}{\eta}\right) \Tilde{v} - (\mu \Tilde{v} - \lambda)\dv{\Tilde{v}}{\lambda}\right]C_\text{sa}\Tilde{\rho},\\
    \dv{\Tilde{\rho}}{\lambda} &= \frac{\Tilde{\rho}}{\gammasa} \left[\frac{k_\rho(\gammasa - 1) + 2 (1 - 1/\eta)}{\mu \Tilde{v} - \lambda} + \frac{1}{\Tilde{P}} \dv{\Tilde{P}}{\lambda}\right].
\end{align}
where 
\begin{equation}
    C_\text{sa} = \frac{\rho_1 v_1 \vs}{P_1} = \frac{(1 - 1/M^2)(\gammasa + 1)}{(\gammasa - 1 + 2/M^2)[1 + (1 - \gammasa) / (2 \gammasa M^2)]}.
\end{equation}

The shocked ambient medium is assumed to have an adiabatic EOS with $\gamma_\text{sa}$. We only consider the case when $\gamma_\text{sa}$ of the compressed ambient medium is the same as $\gamma_\text{sw}$ of the compressed wind.

The numerical integration starts the solution from its known boundary conditions at the ambient shock $\lambda=1^-$, which are $\Tilde{\rho}(1^-)=\Tilde{v}(1^-)=\Tilde{P}(1^-)=1$, and proceeds towards smaller values of $\lambda$ within the shocked ambient region, terminating the integration at the value $\lambda=\lambda_\text{c} \equiv \Rc/\Rs$, which can be detected numerically through its property $\Tilde{v}(\lambda_\text{c}) = \lambda_\text{c} / \mu$. This condition is reached due to the jump of $\Tilde{\rho}$ at contact discontinuity, which causes $\text{d}\Tilde{\rho}/\text{d} \lambda \rightarrow \infty$. As a test of our program, we recomputed the parameter set in Table 3 of KM92b and our results coincided with KM92b and \citet{weaver1977}.

We will use approximate forms of this solution to compute the matching between shocked wind and shocked ambient medium below in Section \ref{sec:matching}.
For that purpose we utilize the same linear approximation for the shocked ambient medium as in KM92b. 
Knowing the expressions for velocity gradients
\begin{align}
    \beta_\text{s} &\equiv \left[\frac{\text{d}\Tilde{v}}{\text{d}\lambda}\right]_{\lambda = 1^-} = \frac{2\mu\gammasa - k_\rho - (1 - 1/\eta)(C_\text{sa}(\mu - 1) - 2)}{C_\text{sa}(\mu - 1)^2 - \gammasa \mu},\\
    \alpha_\text{a} &\equiv \left[\frac{\text{d}\Tilde{v}}{\text{d}\lambda}\right]_{\lambda = \lambda_\text{c}} = \frac{-2(\gammasa + 1) + k_\rho + 2/\eta}{\gammasa \mu},
\end{align}
the linear approximation for velocity based on the average gradient $m = (\beta_\text{s} + \alpha_\text{a}) / 2$, then will be $\Tilde{v}(\lambda) \simeq m (\lambda - 1) + 1$. With $\Tilde{v}(\lambda_\text{c}) = \lambda_\text{c} / \mu \simeq m (\lambda_\text{c} - 1) + 1$, we can get approximate value of $\lambda_\text{c}$,
\begin{equation}\label{eq: lambda_approx}
    \lambda_\text{c} \equiv \frac{\Rc}{\Rs} \simeq \frac{\mu(m - 1)}{\mu m - 1},
\end{equation}
which coincides with Equation (B8a) as derived in KM92b, for large Mach numbers $M \to \infty$.

\subsubsection{Shocked Wind} \label{sec:shocked_wind}
In this section, we pose and solve a set of ODEs for the compressed wind region, extending to the shocked wind region the methods that App.\ B.1 of KM92b utilized for the shocked ambient region. This is in contrast with the approach to the shocked wind in App.\ B.2 of KM92b, which was based on a modified form of the isobaric assumption that had been used by \citet{weaver1977}.

Equation (3.14) in KM92b shows that the position of the wind shock $\Rsw$ can be written as a power-law of time 
$\Rsw(t) \propto t^\varkappa$, where
\begin{equation}\label{eq:varkappa}
\varkappa = \frac{1 + k_\rho + 3 \etain}{2(5 - k_\rho)}\ .
\end{equation}
From now on we restrict the wind to the case $\etain=1$, $k_\rho=2$, giving $\varkappa = 1$.
We introduce the velocity of the wind shock $\vsw(t) = d \Rsw(t) / dt = \varkappa \Rsw(t)/t$ (not the same as $\vsw$ in KM92b, which is defined in the wind frame).
The density of the free wind at the preshock point $r=\Rsw^-$ can be expressed as in KM92b Eq.\  (2.5):
\begin{equation}\label{eq:rhow}
    \rho_\text{w}(t) =
    \frac{\Mdotin(t)}{4\pi R^2_\text{sw}(t) \vin} \left(1 - \frac{\Rsw(t)}{t\vin}\right)^{\etain - 1}=\frac{\Mdotin(t)}{4\pi R^2_\text{sw}(t) \vin}
    \ ,
\end{equation}
where the product $\Mdotin(t)t^{-\etain+1}$ is constant in time,
and applying the choice of a steady wind with $\etain=1$ to simplify the expression for $\rho_\text{w}(t)$.
This also simplifies
$\zeta(t)$ in $\lfrac{d\rho_\text{w}(t)}{dt} = \zeta(t) \rho_\text{w}(t) / t$ to $\zeta(t) = - 2$.
General Rankine-Hugoniot jump conditions \eqref{eq:compression_limit} for the wind shock at $r = \Rsw$ can be written as
\begin{align}
    &\rho_2(t) = \frac{\gammasw + 1}{\gammasw - 1 + 2 /M^2}\rhow(t), \\
    &v_2(t) = \frac{\gammasw - 1 + 2/M^2}{\gammasw + 1} \vin + \frac{2 - 2/M^2}{\gammasw + 1} \vsw(t), \\
    &P_2(t) = \left(\frac{2 - 2/M^2}{\gammasw + 1} + \frac{1}{\gammasw M^2}\right) \rhow(t) (\vsw(t) - \vin)^2.
\end{align}
Conditions for this strong shock approximation are explored in Section \ref{sec:vin} below. 

We define the dimensionless hydrodynamics variables of the self-similar flow\footnote{Note that we use tilde notation ($\Tilde{v}$, $\Tilde{P}$, $\Tilde{\rho}$)  for compressed ambient medium and bar notation ($\bar{v}$, $\bar{P}$, $\bar{\rho}$) for compressed wind.}, by $\xi \equiv \lfrac{r}{\Rsw(t)}$ and $\sigma \equiv \lfrac{v_2}{\vsw}$, such that
$\vin/\vsw=\lfrac{[\sigma(\gamma+1)-2 + 2/M^2]}{(\gamma-1 + 2/M^2)}$,
\begin{equation}
    v(r, t) = v_2(t) \bar{v}(\xi), \quad \rho(r, t) = \rho_2(t) \bar{\rho}(\xi), \quad P(r, t) = P_2(t) \bar{P}(\xi),
\end{equation}
one can obtain the system of ODEs for the shocked wind, by substituting the dimensionless forms into hydrodynamic equations:
\begin{align}
    \sigma \dv{\bar{v}}{\xi} + (\sigma \bar{v} - \xi) \frac{1}{\bar{\rho}} \dv{\bar{\rho}}{\xi} + \frac{2 \sigma \bar{v}}{\xi} + \frac{\zeta}{\varkappa} &= 0,\label{eq:wind_1}\\
    \left(\frac{\gammasw + 1}{2 - 2 /M^2}\right)\sigma(\sigma \bar{v} - \xi) \dv{\bar{v}}{\xi} + \frac{(\vin / \vsw - 1)^2}{C_\text{sw}\bar{\rho}}\dv{\bar{P}}{\xi} + \left(1 - \frac{1}{\varkappa}\right) \bar{v} &= 0,\label{eq:wind_2}\\
    (\sigma \bar{v} - \xi) \left(\frac{-\gammasw}{\bar{\rho}}\dv{\bar{\rho}}{\xi} + \frac{1}{\bar{P}}\dv{\bar{P}}{\xi}\right) + (1 - \gammasw) \frac{\zeta}{\varkappa} - 2 \left(\frac{1 - 1/\varkappa}{\vin / \vsw - 1}\right) &= 0.\label{eq:wind_3}
\end{align}
As expected, replacing $\vin = 0$, and using that $\etain=1$, $\varkappa = 1$, $M \to \infty$ these equations have the same shape as (B5) in KM92b,
after changing the appropriate coefficients and variables.
We rewrite Equations (\ref{eq:wind_1}--\ref{eq:wind_3}) as explicit formulas for the derivatives:
\begin{align}
    \dv{\bar{v}}{\xi} &= \left\{\frac{2\gammasw \sigma \bar{v}}{\xi} + \frac{\zeta}{\varkappa} - \left(\frac{1 - 1/\varkappa}{\vin/\vsw - 1}\right)\left[\frac{C_\text{sw}(\sigma \bar{v} - \xi)}{\vin /\vsw - 1} \frac{\bar{v}\bar{\rho}}{\bar{P}} + 2\right]\right\}\cdot \left[\frac{C_\text{sw}(\gammasw + 1)\sigma}{2 - 2 /M^2} \left(\frac{\sigma \bar{v} - \xi}{\vin/\vsw - 1}\right)^2\frac{\bar{\rho}}{\bar{P}} - \gammasw \sigma\right]^{-1},\label{eq:wind_v}\\
    \dv{\bar{P}}{\xi} &= \left[-\left(1 - \frac{1}{\varkappa}\right)\bar{v} - \left(\frac{\gammasw + 1}{2 - 2/M^2}\right)\sigma(\sigma \bar{v} - \xi)\dv{\bar{v}}{\xi}\right]\frac{C_\text{sw}\bar{\rho}}{(\vin/\vsw - 1)^2},\label{eq:wind_p}\\
    \dv{\bar{\rho}}{\xi} &= \left\{\left[(1 - \gammasw)\frac{\zeta}{\varkappa} -2 \left(\frac{1 - 1/\varkappa}{\vin/\vsw - 1}\right)\right](\sigma \bar{v} -\xi)^{-1} + \frac{1}{\bar{P}}\dv{P}{\xi}\right\}\frac{\bar{\rho}}{\gammasw},\label{eq:wind_rho}
\end{align}
where 
\begin{equation}
    C_\text{sw} = \frac{(\gammasw + 1) (1 - 1/M^2)}{(\gammasw - 1 + 2/M^2)[1 + (1 - \gammasw) / (2\gammasw M^2)]}.
\end{equation}
We can numerically solve the equations above with $\bar{v}(1^+) = \bar{P}(1^+) = \bar{\rho}(1^+) = 1$ as initial conditions, progressing until the contact discontinuity is located through its property $\bar{v}(\xi_c) = \xi_c / \sigma$.
Similarly to KM92b for the compressed ambient region this location of contact discontinuity was chosen due to a singularity appearing for $\text{d} \bar{\rho} / \text{d} \xi$.
These solutions depend on \{$\gamma$, $\vratio$\} as parameters, with the fixed constants $\etain = \varkappa = 1$ and $k_\rho = 2$.

Here we write down gradients of velocity at wind shock and contact discontinuity, which will be used in Section~\ref{sec:matching}:
\begin{align}
    \beta_\text{sw} &\equiv \left[\dv{\bar{v}}{\xi}\right]_{\xi = 1^+} =\left\{2\gammasw\sigma + \frac{\zeta}{\varkappa} - \left(\frac{1 - 1/\varkappa}{\vin /\vsw - 1}\right)\left[\frac{C_\text{sw} (\sigma - 1)}{\vin/\vsw - 1} + 2\right]\right\} \cdot \left[\frac{C_\text{sw}(\gammasw + 1) \sigma}{2 - 2 /M^2}\left(\frac{\sigma - 1}{\vin /\vsw - 1}\right)^2 - \gammasw \sigma\right]^{-1},\label{eq:grad_v_sw}\\
    \alpha_\text{w} &\equiv \left[\dv{\bar{v}}{\xi}\right]_{\xi = \xi_\text{c}} = \frac{-2(\gammasw - (\vin / \vsw - 1)^{-1}) - (\zeta + 2 (\vin / \vsw - 1)^{-1}) / \varkappa}{\gammasw \sigma}\label{eq:grad_cw}.
\end{align}

We notice that the solution in the shocked wind depends on the parameter $\vratio$, which is not known initially.
In section \ref{sec:matching}, we provide a viable method to solve for this parameter and obtain a complete solution by matching with the shocked ambient solution at the contact discontinuity.

\subsubsection{Matching Shocked Wind and Ambient Medium} \label{sec:matching}

In this section, we connect the two solutions found in Section \ref{sec:method}, matching them to obtain results. To demonstrate the purpose, approximate solutions for the shocked ambient medium are presented, and a new approximate solution for the shocked wind region is constructed. A solution based on the exact ODEs can also be utilized for matching, leading to a shooting method instead of a closed-form equation.

Systems (B5) of KM92b, and Equations (\ref{eq:wind_v}-\ref{eq:wind_rho}) could be solved independently if the \{$\gamma$, $\vratio$\} parameters were known initially, but, as we mention in the Section \ref{sec:shocked_wind}, $\vratio$ is initially unknown. The value of this parameter can be found by matching and connecting the solutions of the shocked wind and ambient medium at the contact discontinuity, across which both velocity and pressure must be continuous. To match solutions of the shocked wind and shocked ambient medium, we chose the following approach: we analytically approximate the solutions for (B5) of KM92b and (\ref{eq:wind_v}-\ref{eq:wind_rho}), then solve the matching problem for these analytical approximations. A more accurate solution can be found by a shooting method based on the exact ODEs.

The approximate solutions for the compressed ambient follow the same linear approximations as in KM92b, as in Equation \eqref{eq: lambda_approx} above.
For the compressed wind the approximation needs a different approach.
We utilize the gradients in Equations \eqref{eq:grad_v_sw}--\eqref{eq:grad_cw} to obtain approximate form for the solutions of the compressed wind.
For the shocked wind solution
the linear approximation is not appropriate because the curvature is significant. We adopt a hyperbola as an approximation to the ODE solution $\bar{v}(\xi)$, a curve of a similar shape.\footnote{We have also considered a parabola approximation, but the results were not an improvement over the linear approximation.}
We proceed with this hyperbola shape with a procedure similar to the one used for the approximate shocked ambient medium. Knowing the velocity gradient expressions (\ref{eq:grad_v_sw} and \ref{eq:grad_cw}), we let $n = (\beta_\text{sw} + \alpha_\text{w}) / 2$, then the hyperbola approximation for velocity will be $\bar{v}(\xi) \simeq -n (1/\xi - 1) + 1$, a curve passing through the point $\bar{v}(\xi=1)=1$. As we can see both $\beta_\text{sw}$ and $\alpha_\text{w}$ are functions of $\vratio$. We find the value of that parameter by using the continuity of velocity and pressure at the contact discontinuity. For velocity, we get
\begin{equation}
    \xi_\text{c} \vsw = \lambda_\text{c} v_\text{s},
\end{equation}
and for pressure
\begin{equation}
    \frac{\bar{P}(\xi_\text{c})}{\Tilde{P}(\lambda_\text{c})} \frac{\vsw^2}{\vs^2}\left(\sigma - \vratio\right)^2 = \delta^{-1} \frac{\lambda_\text{c}^{k_\rho}}{\xi_\text{c}^2} \Rc^{2-k_\rho}\ , 
\end{equation}
where $\delta = \rho_\text{0w}/\rho_\text{0a}$ is the ratio of free density power laws that we can assume is known initially. After substituting one into another we obtain the equation that we solve for $\vratio$:
\begin{equation}\label{eq: v_in2sw}
    \frac{\bar{P}(\xi_\text{c})}{\Tilde{P}(\lambda_\text{c})} \left(\sigma - \vratio\right)^2 = \delta^{-1},
\end{equation}
using $k_\rho = 2$.
To solve \eqref{eq: v_in2sw}, we use linear approximations for $\bar{P}(\xi)$ and $\Tilde{P}(\lambda)$. The corresponding expressions for gradients of pressure:
\begin{align}
    \chi_\text{s} &\equiv \left[\dv{\Tilde{P}}{\lambda}\right]_{\lambda = 1^-} = \left[-\left(1 - \frac{1}{\eta}\right) - (\mu - 1) \beta_\text{s}\right]C_\text{sa}, \\
    \chi_\text{sw} &\equiv \left[\dv{\bar{P}}{\xi}\right]_{\xi = 1^+} = \left[-\left(1 - \frac{1}{\varkappa}\right)
    - \frac{(\gammasw + 1)\sigma(\sigma - 1)}{2 - 2 / M^2} \beta_\text{sw}\right] \frac{C_\text{sw}}{(\vratio - 1)^2}.
\end{align}
Then linear approximations for pressure $\Tilde{P}(\lambda) \simeq \chi_\text{s}(\lambda - 1) + 1$, $\bar{P}(\xi) \simeq \chi_\text{sw}(\xi - 1) + 1$. To calculate $\Tilde{P}(\lambda_\text{c})$ we use the linear approximation and \eqref{eq: lambda_approx} formula, for $\bar{P}(\xi_\text{c})$ it is a bit more tricky. For every value of $\vratio$ we can solve $\bar{v}(\xi_\text{c}) = \xi_\text{c}/\sigma \simeq -n (1/\xi_\text{c} - 1) + 1$, which transposes to solving the cubic equation 
\begin{equation}\label{eq: xi_approx}
    \frac{\alpha_\text{w}}{2} \xi^3 - \left(\alpha_\text{w} + \frac{1}{\sigma}\right)\xi^2 + \left(\frac{\beta_\text{sw}}{2} + 1\right)\xi - \frac{\beta_\text{sw}}{2} = 0, \text{ where }\alpha_\text{w}, \ \beta_\text{sw}, \ \sigma \text{ are functions of } \vratio.
\end{equation}
The root of \eqref{eq: xi_approx} is the approximate value of $\xi_\text{c}$ as a function of $\vratio$, which could be substituted into calculating approximate value of $\chi_\text{sw}$ also as a function of $\vratio$.

We have the following equation for $\vratio$ 
\begin{equation}\label{eq:final}
    \frac{\chi_\text{sw}(\xi_\text{c} - 1) + 1}{\chi_\text{s}(\lambda_\text{c} - 1) + 1} \left(\sigma - \vratio\right)^2 = \delta^{-1}, \text{ where }\chi_\text{sw}, \ \xi_\text{c}, \ \sigma \text{ are depending on }\vratio,
\end{equation}
that could be solved with Newton–Raphson method with expressions (\ref{eq: lambda_approx}, \ref{eq: xi_approx}) approximating $\lambda_\text{c}$ and $\xi_\text{c}$. After finding the approximate value of $\vratio$, we reverse substitute it into \eqref{eq: xi_approx} and get the approximate value of $\xi_\text{c}$.

\subsubsection{Matched Solution}
\label{sec:matched_solution}

\begin{figure}[ht]
    \centering
    \includegraphics[width=\linewidth]{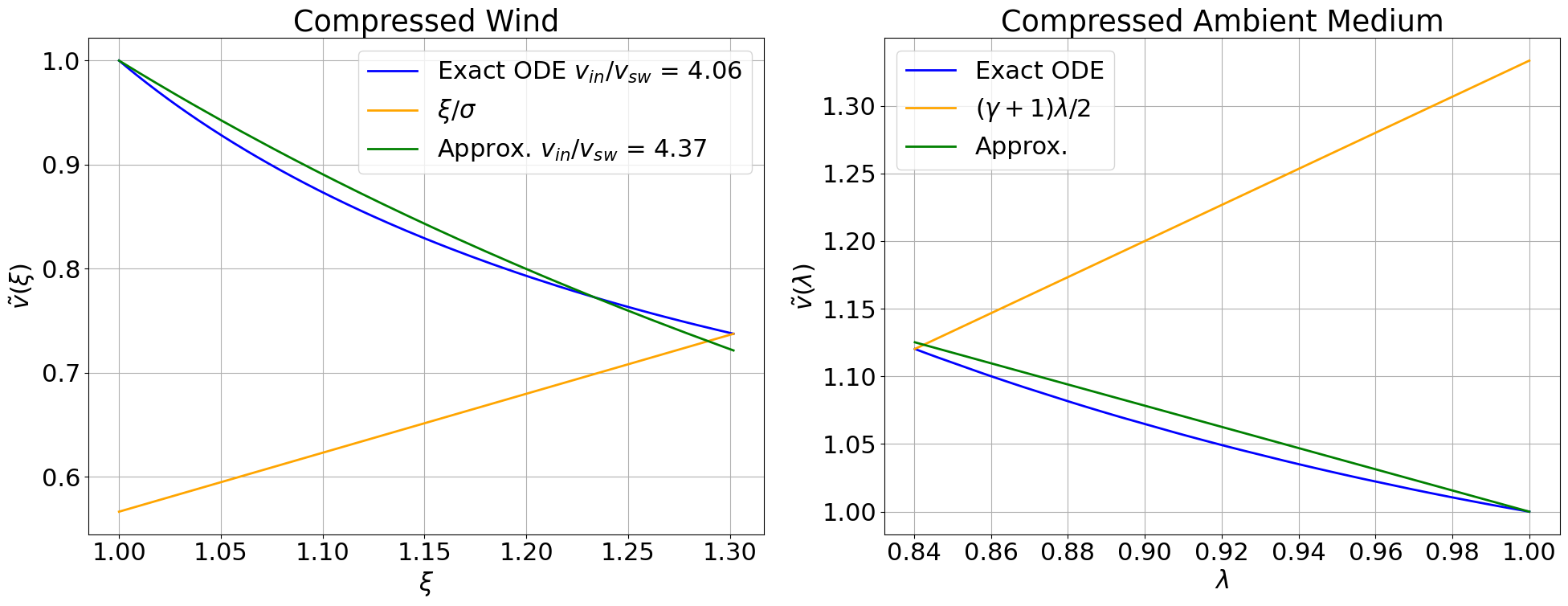}
    \caption{Exact and approximate solutions for $\gamma = 5/3$, $k_\rho = 2$, $\eta_\text{in} = 1$, $\delta = 0.1$ with strong shock assumption (hence $\mu \to 2 / (\gammasa +1)$). Blue lines show solutions of the exact ODEs (Eqs.\ \ref{eq:wind_v}--\ref{eq:wind_rho} in the compressed wind, and B5 of KM92b in the compressed ambient medium). Green lines show approximations (hyperbola for compressed wind and linear for compressed ambient medium). Value of $\vratio = 4.06$ for the exact solution of compressed wind was obtained through \textsc{ZeusTW} simulations for the same set of parameters. The intersection of orange lines with blue and green curves shows the exact and approximate values for $\lambda_\text{c}$ and $\xi_\text{c}$ respectively.}
    \label{fig:approxes}
\end{figure}

\begin{figure}[ht]
    \centering
    \includegraphics[width=0.6\linewidth]{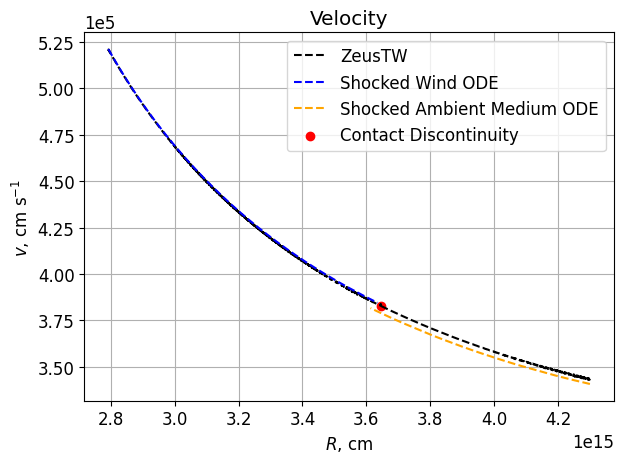}
    \caption{Comparison of velocity profiles obtained via exact ODE solutions with strong shock assumption for compressed wind (Eqs.\ \ref{eq:wind_v}--\ref{eq:wind_rho}, blue) and compressed ambient medium (B5 of KM92b, orange) to numerical \textsc{ZeusTW} simulation (black) in the bubble shell, using $\vin=12\kms$.}
    \label{fig:compare} 
\end{figure}

Using the method described in the previous subsection, we obtained the exact ODE solutions and analytical approximations for the case $\gamma = 5/3$, $k_\rho = 2$, $\eta_\text{in} = 1$, $\delta = 0.1$ (Fig.\ \ref{fig:approxes}). 
Additionally, we compared the exact ODE solutions to the \textsc{ZeusTW} numerical simulation (Fig.\ \ref{fig:compare}).
In the future, we can also compare them using Athena++ \citep{stone2020}, and our in-house codes SADHANA \citep{bandopadhyay2022}, Astaroth \citep{vaisala2023}, and Kinetic Modules (KM, \citealt{motoyama2015, motoyama2024}).

These figures show great compliance both the exact ODE solutions and the approximate hyperbolic solutions (Fig.\ \ref{fig:approxes}) and exact ODE solutions with \textsc{ZeusTW} simulations (Fig.\ \ref{fig:compare}), they also show the velocity structure of the bubble.\footnote{The slight mismatch near the contact discontinuity does not represent a jump but merely a minor imperfection in the matching between ambient and wind portions of the solution.} Table \ref{tab:shell_thickness} shows exact and approximate results, including the dimensionless thickness of the shell as the ratio
\begin{equation}
    \frac{\Rs}{\Rsw}=\frac{\xi_\text{c}}{\lambda_\text{c}}
\end{equation}
derived from the values of $\lambda_\text{c}$ and $\xi_\text{c}$. 
\begin{table}[ht]
\centering
\begin{tabular}{*{5}c}
\hline\hline
Parameter & $\lambda_\text{c}$  & $\vratio$ & $\xi_\text{c}$ & $\Rs / \Rsw$ \\ \hline
Exact & 0.840 & 4.06 & 1.302 & 1.55\\
Approximate & 0.843 & 4.37 & 1.289 & 1.53 \\ \hline
\end{tabular}
\caption{For $\gamma = 5/3$, $k_\rho = 2$, $\eta_\text{in} = 1$, $\delta = 0.1$ under the strong shock assumption.}
\label{tab:shell_thickness}
\end{table}
For the solution found in Table \ref{tab:shell_thickness}, the value of $\lambda_c$ is independent of the parameter $\delta$ of density ratio, but other variables such as $\xi_c$ and the total thickness can depend on $\delta$.

The numbers in Table \ref{tab:shell_thickness} and the velocity profiles in Figure \ref{fig:approxes} are a proof of concept, demonstrating the ability of this approximate analytical method to self-consistently and separately find the relative thickness of the spherical bubble, without needing to simulate a PDE, assuming constant pressure, and with the solution of the ODE replaced by simple linear and hyperbola approximations. A numerical solution based on the exact ODEs with shooting is also possible. The thickness of a bubble is an important characteristic, decisive in its properties of stability \citep{shang2020}, cooling (KM92b), and observable features \citep{shang2023_signatures} in both the spatial and the spectral spaces.

\begin{figure}
    \centering
    \includegraphics[width=\linewidth]{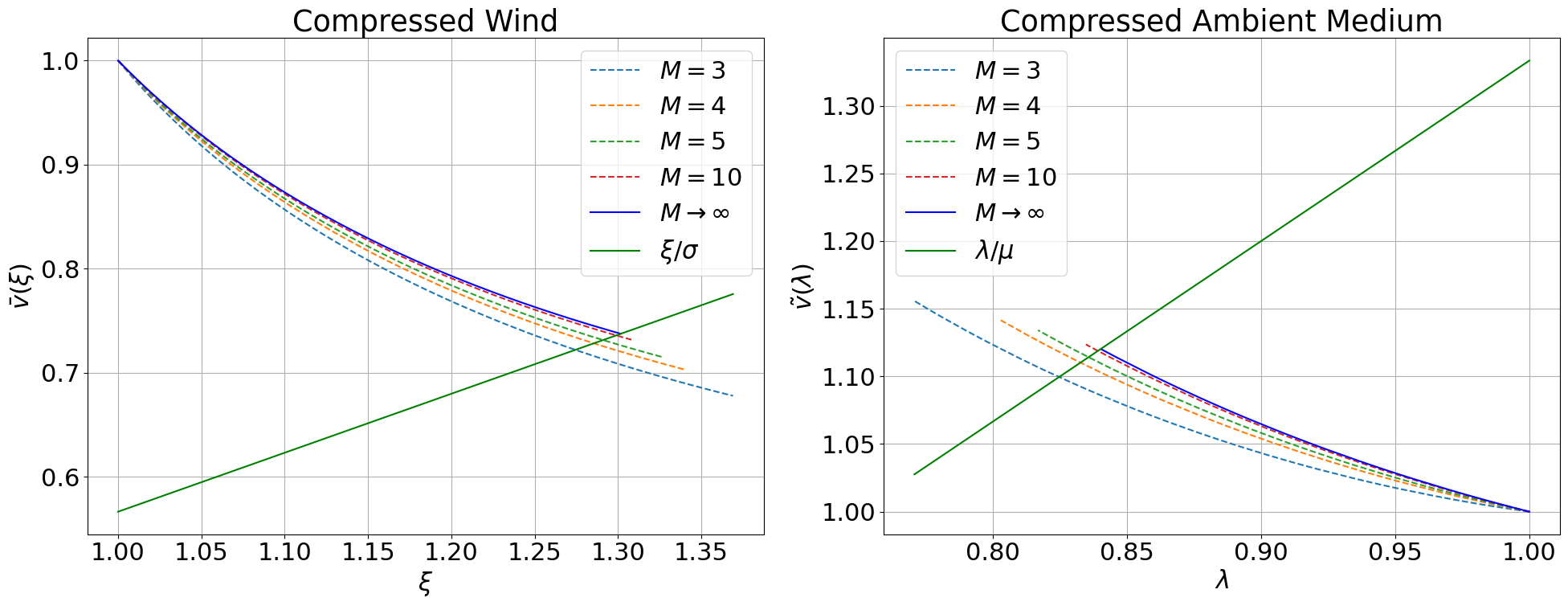}    \includegraphics[width=\linewidth]{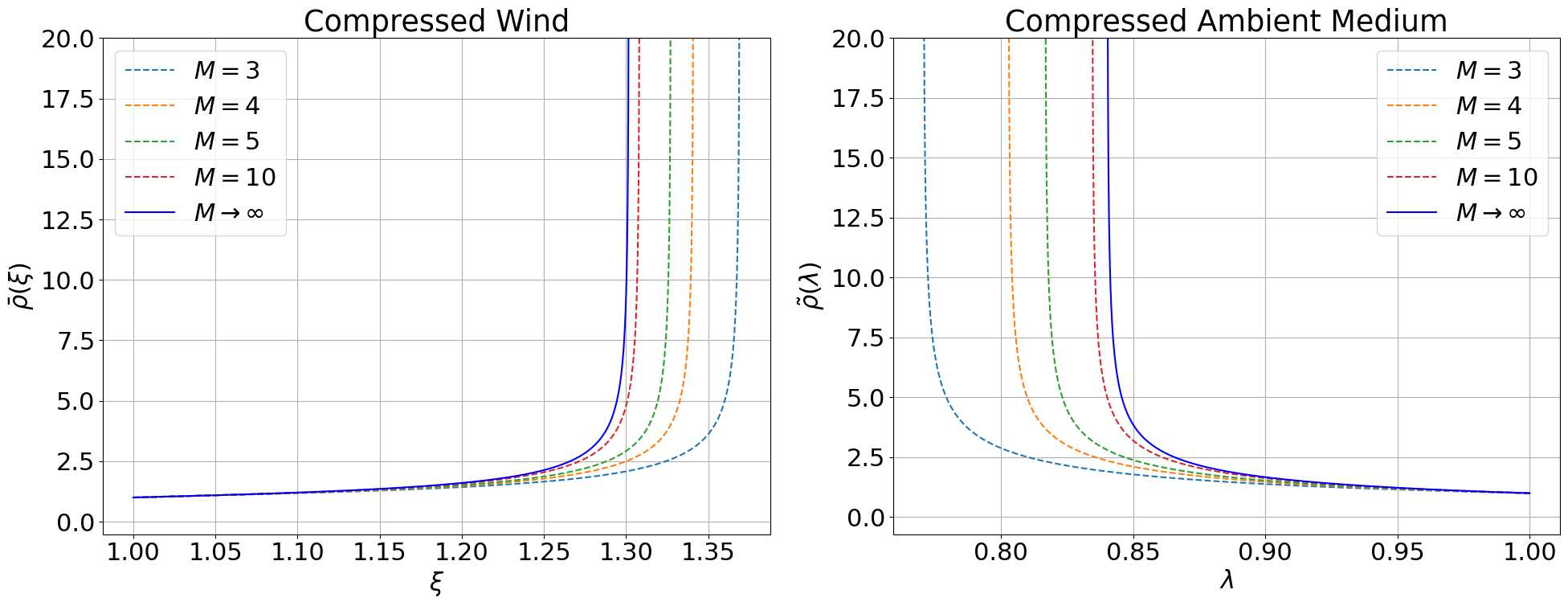}
    \includegraphics[width=\linewidth]{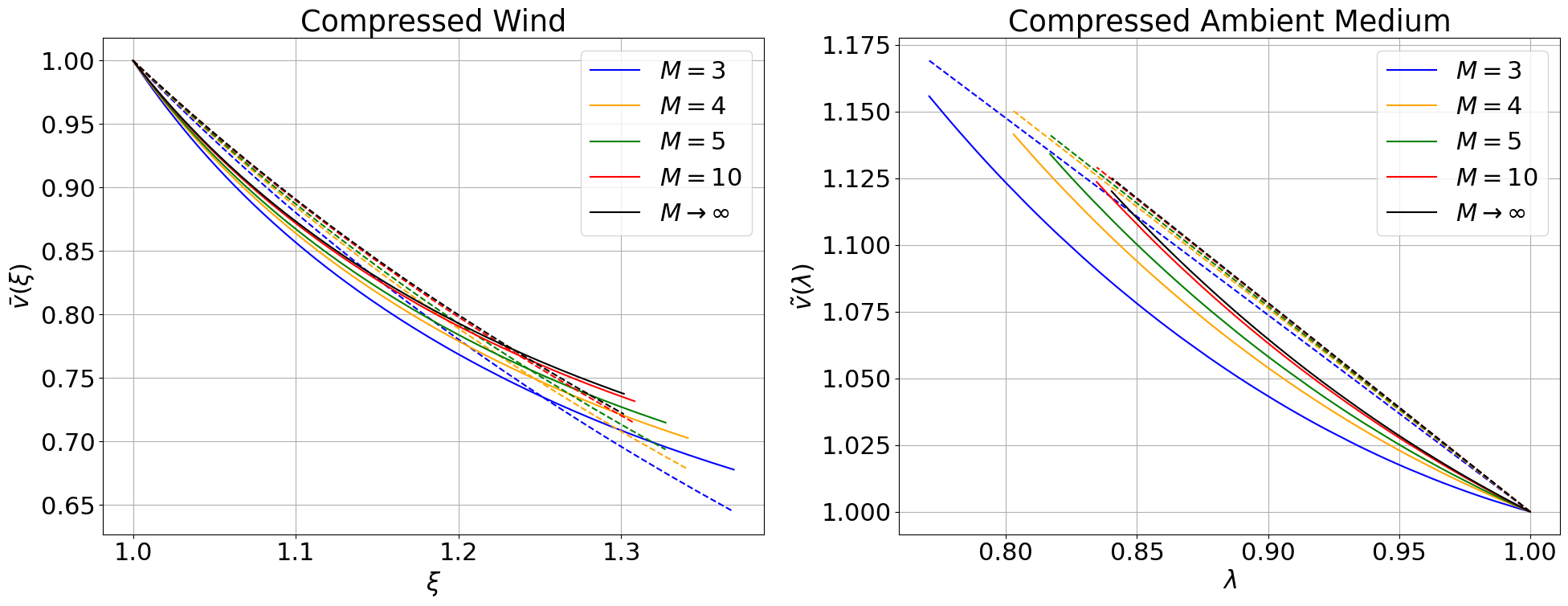}
    \caption{Investigating solution dependence on the Mach number $M=M_\text{sw}=M_\text{sa}$ (see Section \ref{sec:matched_solution}). Top rows: velocities and densities for the exact ODE solutions in the two compressed regions, using for the compressed wind $\vin/\vsw = 4.06$. The lines $\xi/\sigma$ and $\lambda/\mu$ in the top row have been drawn for the case $M=\infty$ ($\sigma$ and $\mu$ depend on Mach number). Bottom row: comparison of exact ODE (solid lines) to the hyperbola and linear approximations (dashed lines, computed using the $\vin/\vsw$ values obtained through matchings shown in the $M_\text{sw}=M_\text{sa}$ rows of Table \ref{tab:shell_thickness_mach}).}
    \label{fig:mach_vel_rho}
\end{figure}

\begin{figure}
    \centering
    \includegraphics[width=\linewidth]{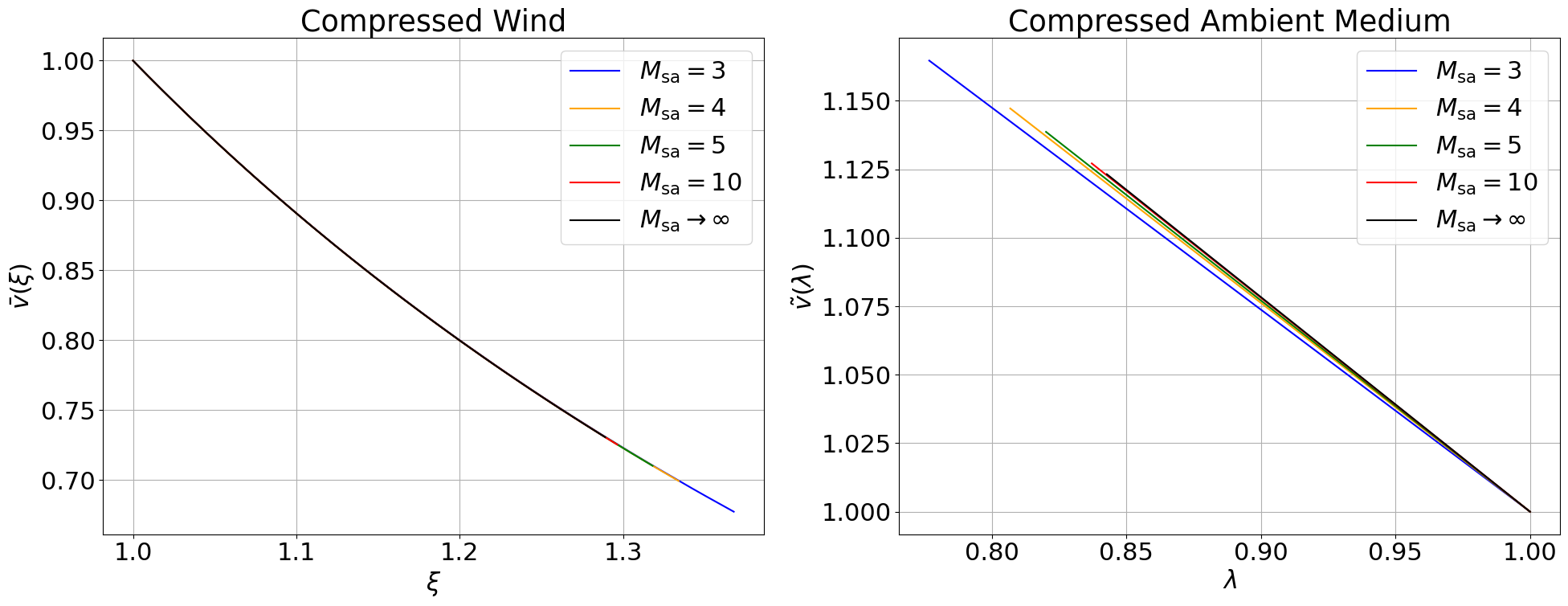}
    \caption{Hyperbolic and linear approximations using values for $\vin/\vsw$ obtained through matching. Computed for $M_\text{sw} =\infty$, and five values of $M_\text{sa}$ in the range from 3 to $\infty$ (Table \ref{tab:shell_thickness_mach}).}
    \label{fig:mach_ambient}
\end{figure}

Additionally, we investigated how exact ODE and approximate solutions depend on the Mach numbers $M_\text{sa}$ and $M_\text{sw}$ of the ambient and wind sides. Figure \ref{fig:mach_vel_rho} explores cases with $M_\text{sa}=M_\text{sw}$, and Figure \ref{fig:mach_ambient} explores the approximate solutions for cases with $M_\text{sw}=\infty$.
Parameters and thickness results of the approximate solutions are shown in Table \ref{tab:shell_thickness_mach}. Matchings were fully computed for all cases based on the hyperbola approximation. The values of $\vratio$ in column 4 of Table \ref{tab:shell_thickness_mach} have a very narrow range ($\sim1\%$), allowing in Figure \ref{fig:mach_vel_rho} to trace the corresponding ODE solutions using $\vratio=4.06$.
The top two rows of Figure \ref{fig:mach_vel_rho} show velocity and density in ODE solutions.
As we can see, for $M = 10$ the error becomes negligible ($|\xi_\text{c}^M - \xi_\text{c}^\infty|/ \xi_\text{c}^M < 0.01$, $|\lambda_\text{c}^M - \lambda_\text{c}^\infty|/ \lambda_\text{c}^M < 0.01$),  for smaller $M = 3$ the error is noticeable ($|\xi_\text{c}^M - \xi_\text{c}^\infty|/ \xi_\text{c}^M \simeq 0.05$, $|\lambda_\text{c}^M - \lambda_\text{c}^\infty|/ \lambda_\text{c}^M \simeq 0.09$), but the approximation still gives decent accuracy.
Figure \ref{fig:mach_ambient} and the bottom row of Figure \ref{fig:mach_vel_rho} show the results of the linear and hyperbola approximation when varying the Mach numbers.
We obtained maximum error for $\lambda^{*}_\text{c}$ with $M_\text{sa} = M_\text{sw}$, $|\lambda_\text{c}^{*} - \lambda_\text{c}|/ \lambda_\text{c} < 0.01$ and for $\xi^{*}_\text{c}$, $|\xi_\text{c}^{*} - \xi_\text{c}|/ \xi_\text{c} \lesssim 0.03$, here $*$ is for approximate result. For $M_\text{sw} = \infty$ and finite $M_\text{sa}$ maximum error for $\lambda^{*}_\text{c}$ is the same, for $\xi^{*}_\text{c}$ it increases to $\sim 0.06$.

\begin{table}[ht]
\centering
\begin{tabular}{*{7}c}
\hline\hline
$M_\text{sa}$ & $M_\text{sw}$ & $\lambda_\text{c}$  & $\vratio$ & $\xi_\text{c}$ & $\Rs / \Rsw$ \\ \hline
3 & $\infty$ & 0.777 & 4.34 & 1.283 & 1.65\\
4 & $\infty$ & 0.807 & 4.36 & 1.285 & 1.59\\
5 & $\infty$ & 0.820 & 4.36 & 1.285 & 1.57\\
10 & $\infty$ & 0.837 & 4.37 & 1.285 & 1.54\\\hline
3 & 3 & 0.777 & 4.43 & 1.325 & 1.71\\
4 & 4 & 0.807 & 4.41 & 1.311 & 1.62\\
5 & 5 & 0.820 & 4.40 & 1.302 & 1.59\\
10 & 10 & 0.837 & 4.38 & 1.290 & 1.54\\ \hline
\end{tabular}
\caption{Approximate solutions for $\gamma = 5/3$, $k_\rho = 2$, $\eta_\text{in} = 1$, $\delta = 0.1$.}
\label{tab:shell_thickness_mach}
\end{table}

\subsection{Numerical method: Dependence on wind velocity}
\label{sec:vin}

In this section, we explore the dependence of the solutions on the wind velocity $\vin$, focusing on its effects on the thickness ratio $\Rs/\Rsw$. We have computed a set of high-resolution 1D numerical simulations of the spherical problem exploring different values of $\vin$ and $\gamma$.
The results largely agree with the methods of Sections \ref{sec:method}. However, in specific ranges of small $\gamma$ and $\vin$ values within the set (with compression factors limited by Mach number), thickness dependence on $\vin$ has been encountered.
These dependence ranges include the (local) isothermal limit $\gamma=1.00001$, analogous to the study in Appendix A of \citet{shang2020} (hereafter, S20A).

\begin{figure}[ht]
    \centering
    \includegraphics[height=2.4in]{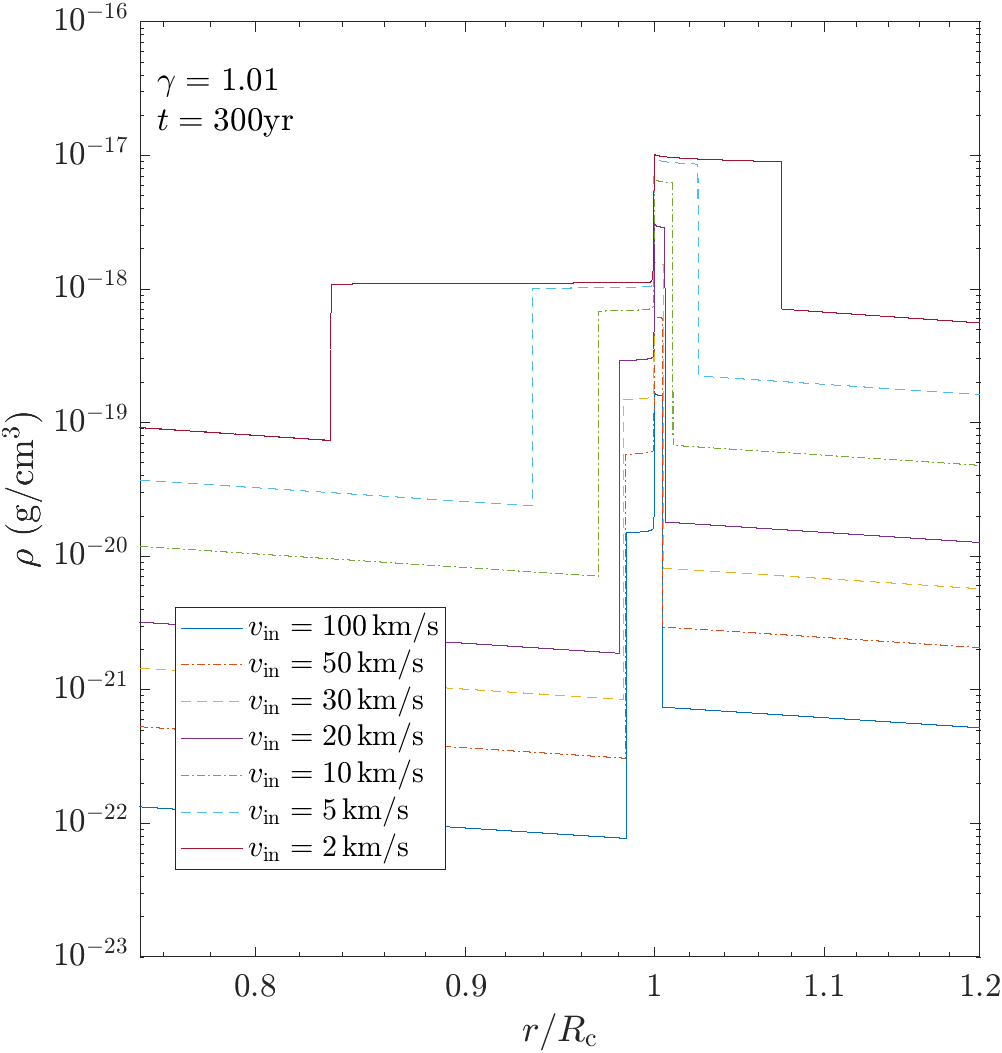}
    \includegraphics[height=2.4in]{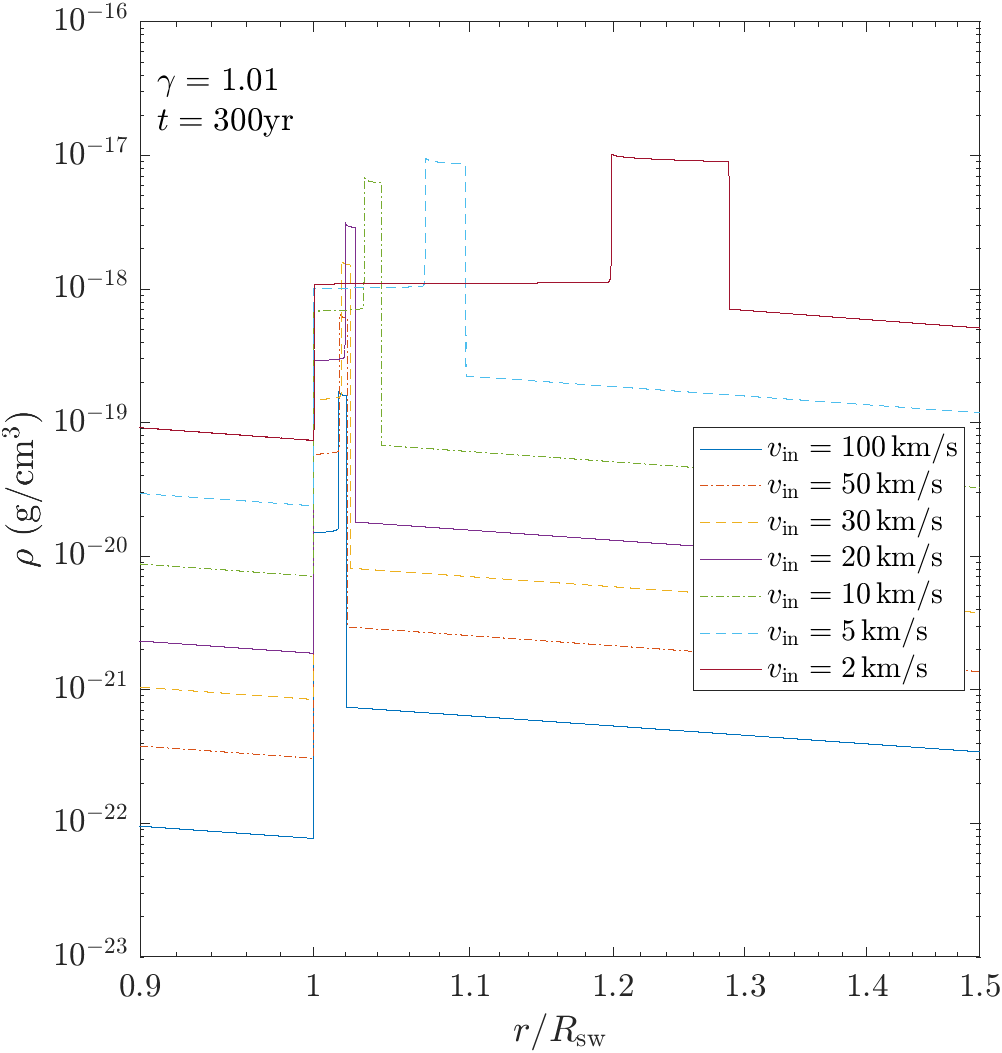}
    \caption{Gas density $\rho$ at $t=300\yr$ for the small value of $\gamma=1.01$. Position coordinate $r$ scaled respectively with
    $\Rc$ and $\Rsw$.}
    \label{fig:vindep1}
\end{figure}

\begin{figure}[ht]
    \centering
    \includegraphics[height=2.4in]{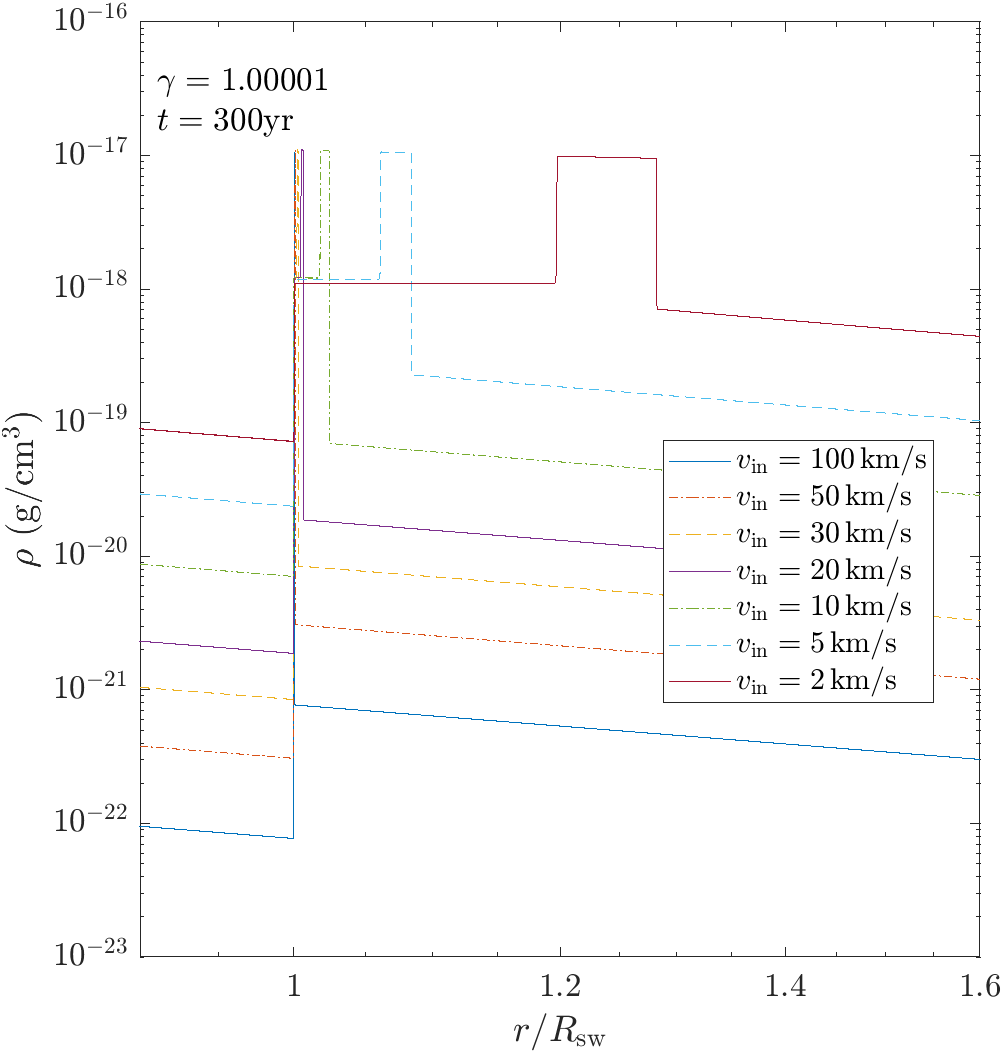}
    \includegraphics[height=2.4in]{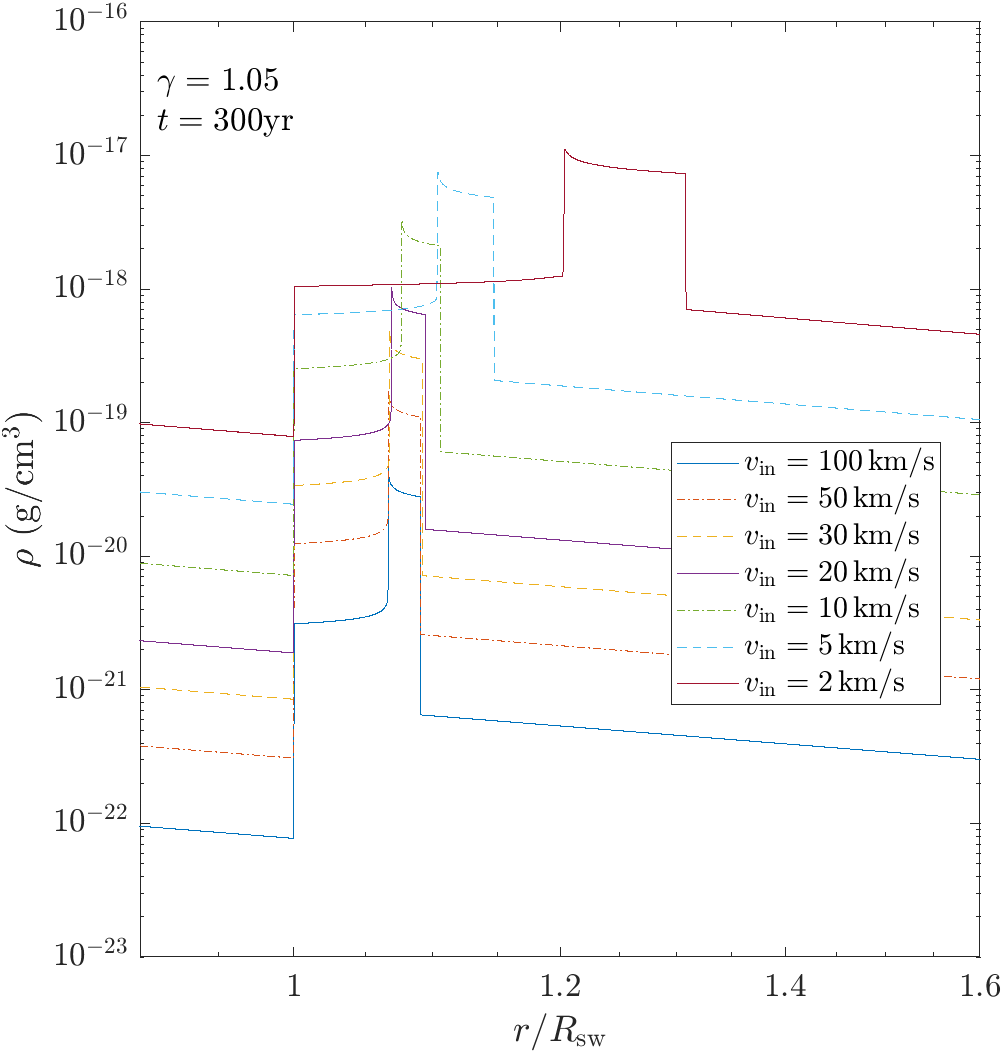}
    \includegraphics[height=2.4in]{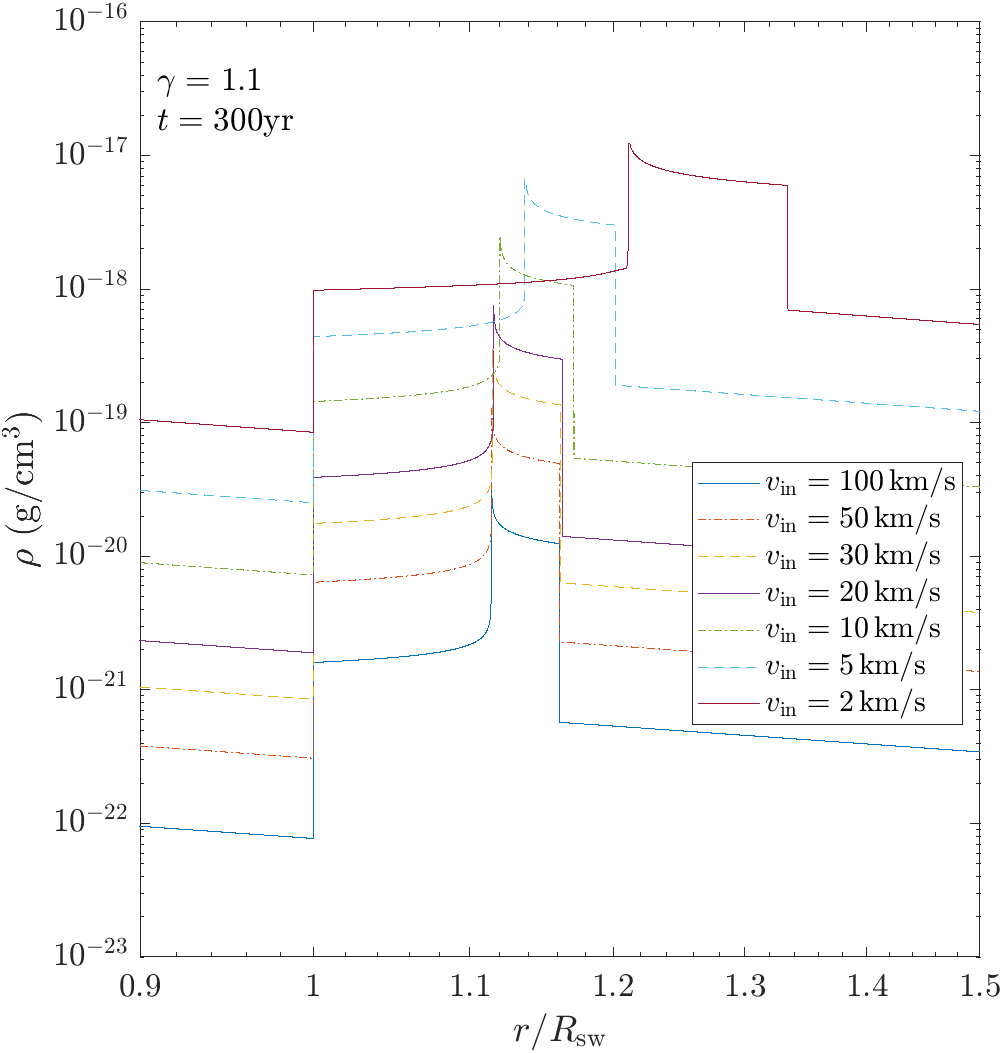}
     \caption{Gas density $\rho(r/\Rsw)$ at $t=300\yr$ for the small values of $\gamma=1.00001$, $1.05$, and $1.1$. }
     \label{fig:vindep2}
\end{figure}

\begin{figure}[ht]
    \centering     
    \includegraphics[height=2.4in]{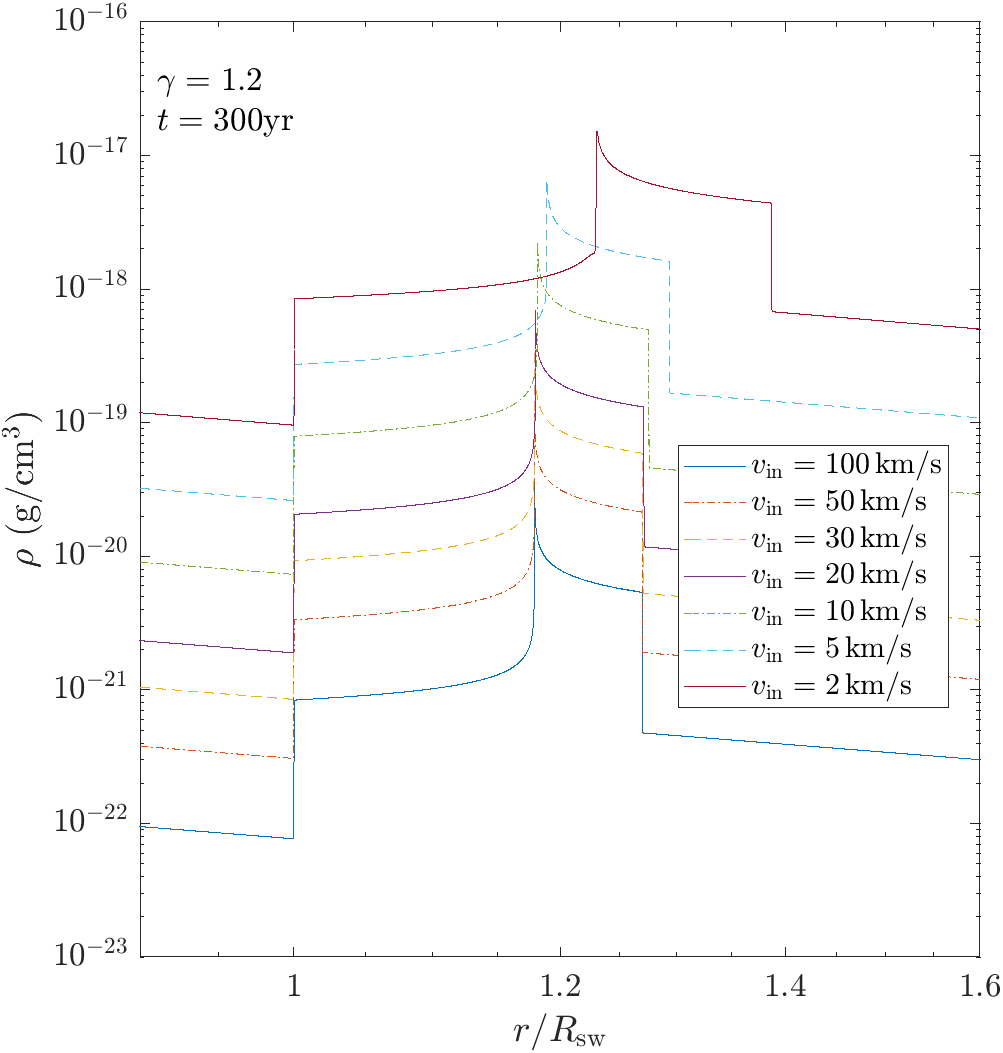}
    \includegraphics[height=2.4in]{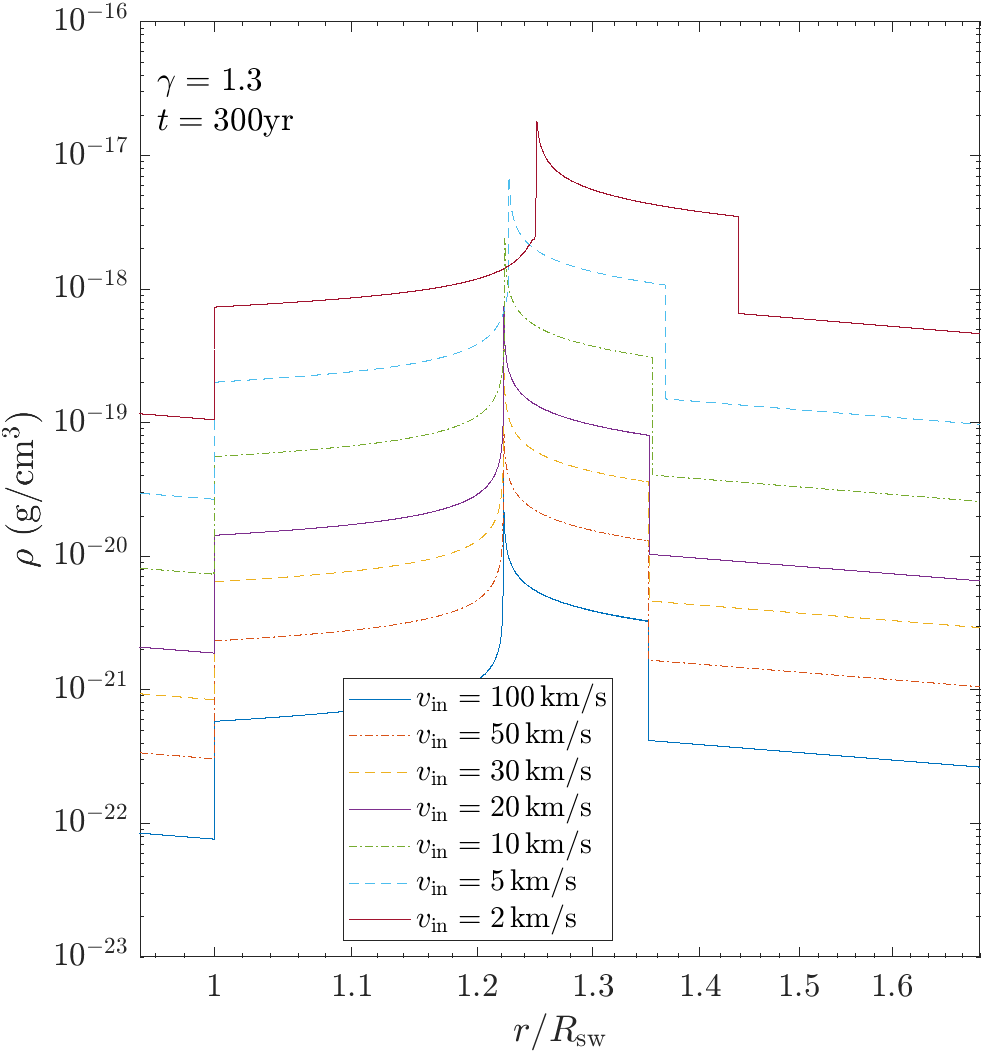}
    \includegraphics[height=2.4in]{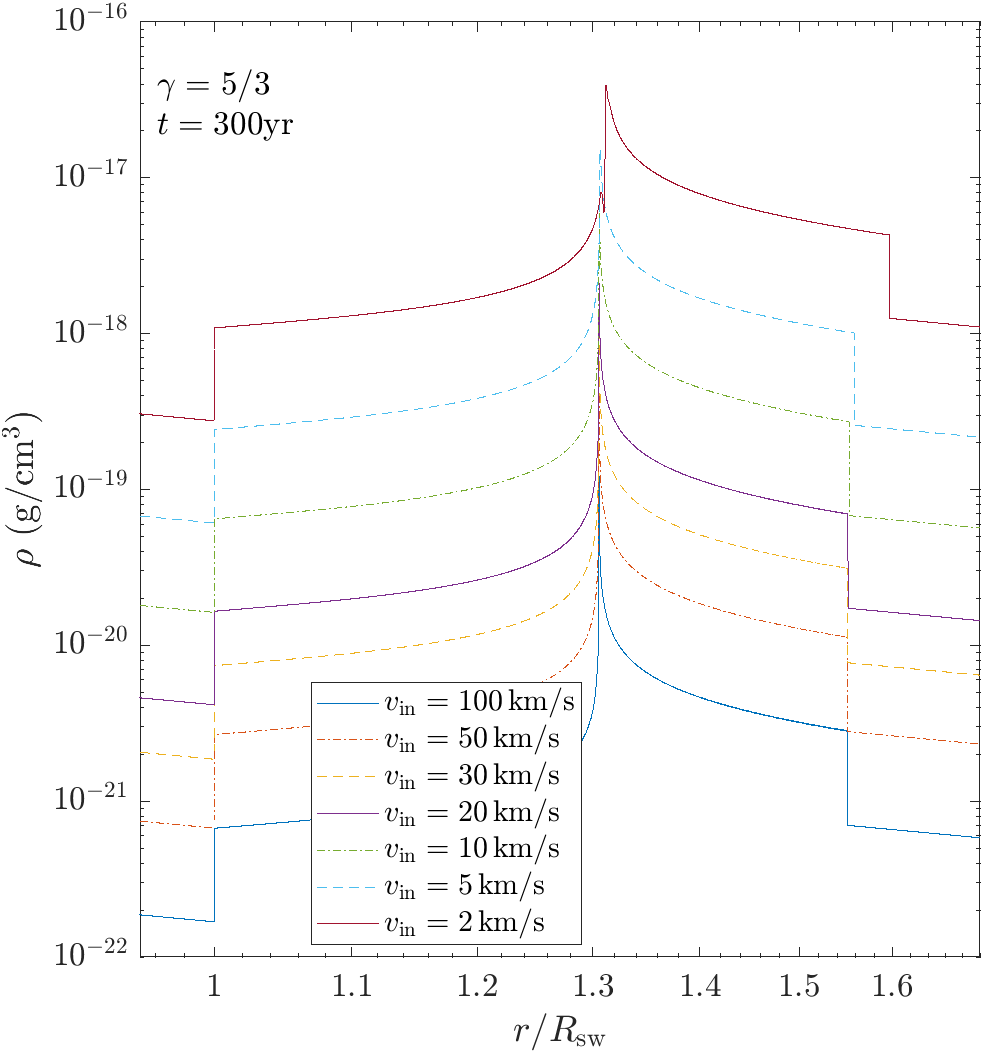}
    \caption{Same as Fig.\ \ref{fig:vindep2} for moderate and large values of polytropic $\gamma$.}
      \label{fig:vindep3}
\end{figure}

In this section we show results for a range of $\gamma=5/3$ to $1.00001$, going from monoatomic adiabatic to nearly isothermal, and $\vin=100$ to $2\kms$, to be compared to the ambient sound speed $\aamb=0.2\kms$, and to the wind sound speed ${\awind}_0=0.6\kms$ at the wind inlet $r=\rin=1.5\au$. (Due to adiabatic cooling, the value of $\awind$ at the preshock point $r={\Rs}^-$ is much smaller than ${\awind}_0$.) The wind is constant in time ($\etain=1$). We set $k_\rho=2$, and therefore $\eta=1$, The wind/ambient density ratio has been kept to 0.1, as in Section \ref{sec:matched_solution}.
The numerical simulation code does not a priori assume that shock is strong.
Figures \ref{fig:vindep1}--\ref{fig:vindep3} show the resulting density profiles as functions of $r/\Rsw$ with $r$ ranging from $<\Rsw$ to $>\Rs$. Results are reported at run time $t=300\yr$ of the simulation, sufficient to achieve an essentially self-similar state within the shell while still unaffected by the outer boundary of the domain.

\begin{figure}[ht]
    \centering
    \includegraphics[height=2.4in]{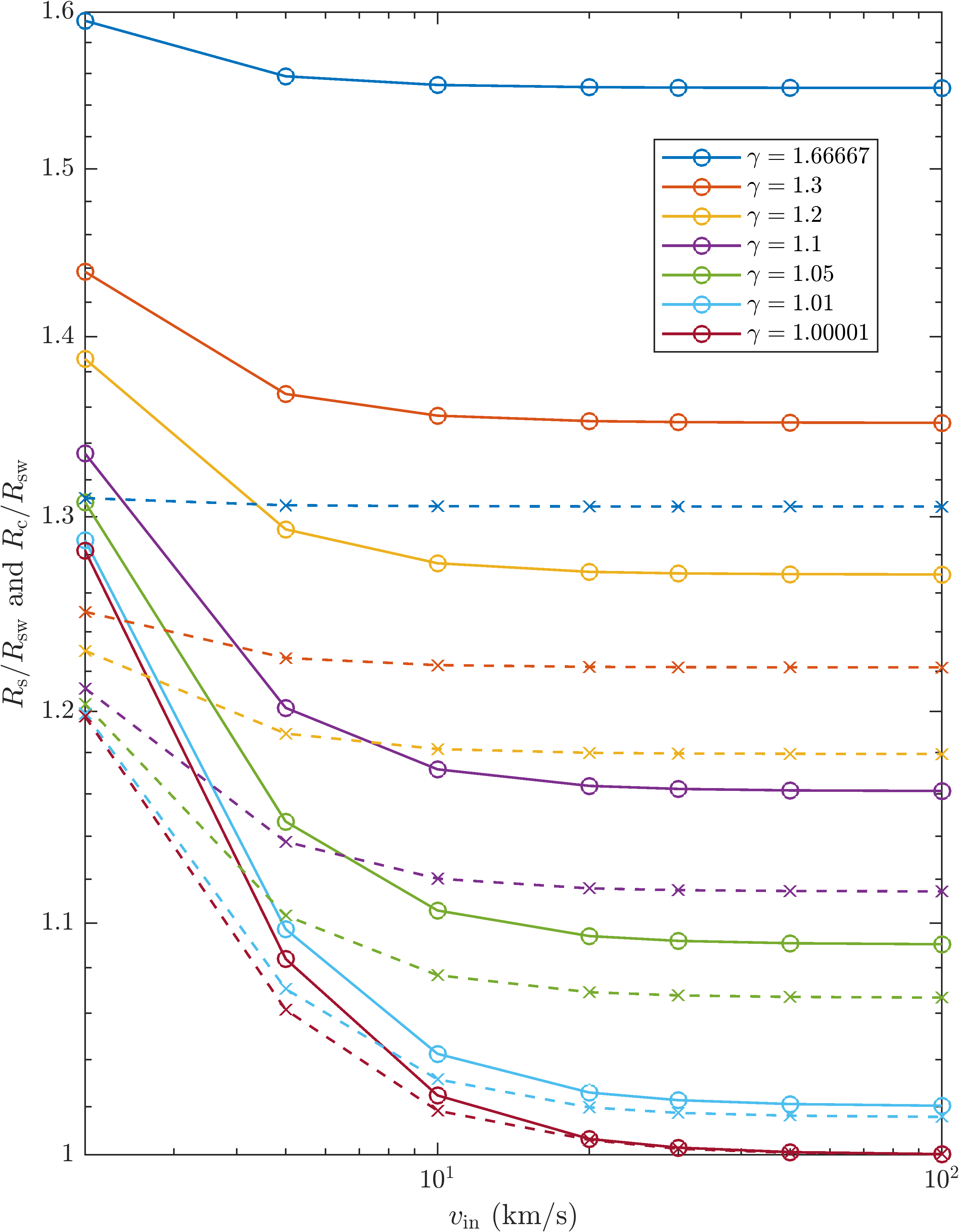}
    \includegraphics[height=2.4in]{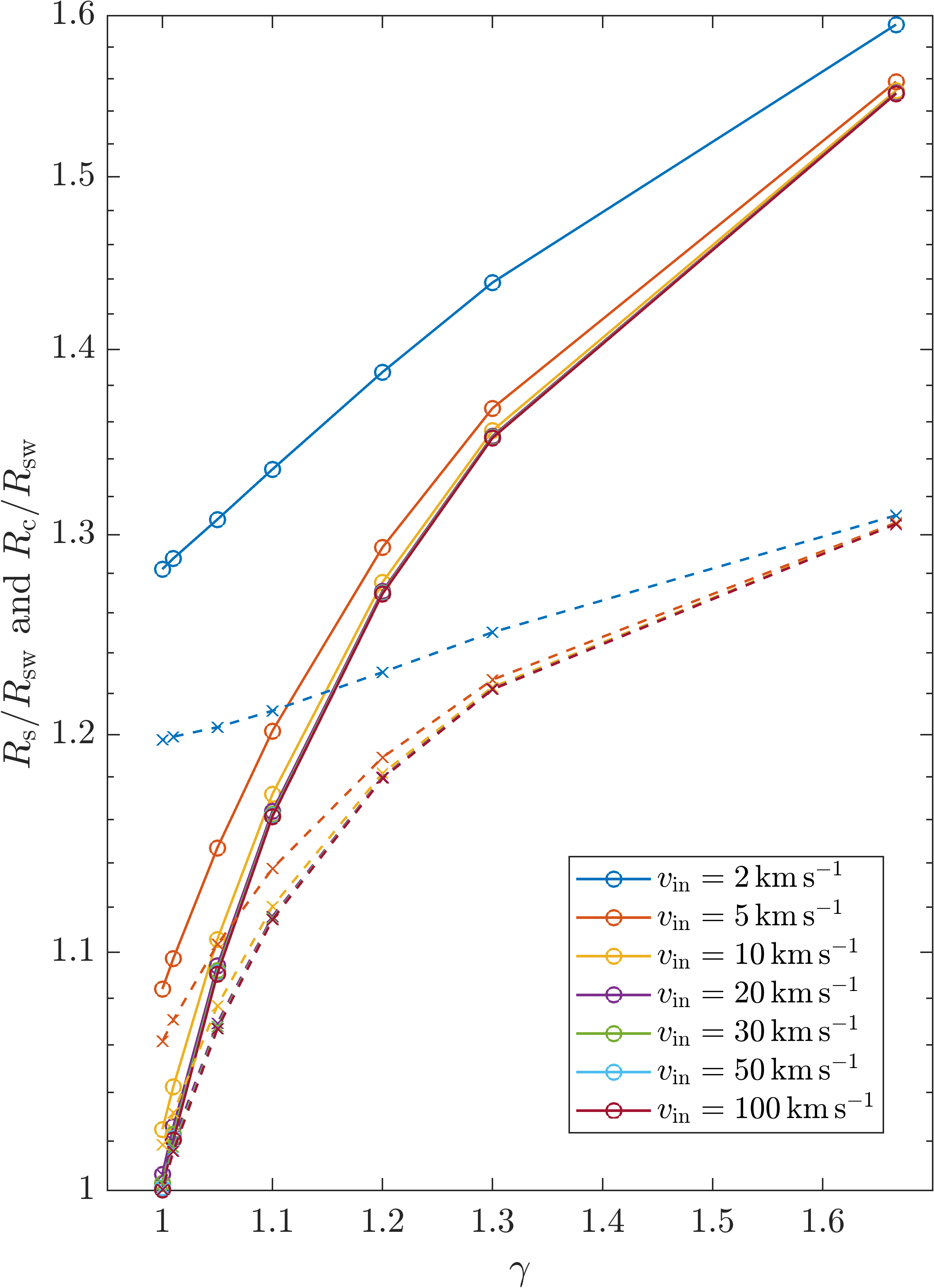}
    \caption{Thickness ratios $\Rs/\Rsw$ (solid lines) and $\Rc/\Rsw$ (dashed lines) for the set of $7\times7$ values of $\vin$ and $\gamma$, for sound speeds $\awind=3\aamb=0.6\kms$ and a background density ratio $(r^2\rhow)/(r^2\rhoa)=0.1$.}
      \label{fig:thickness}
\end{figure}

\begin{figure}[ht]
    \centering
    \includegraphics[height=2.4in]{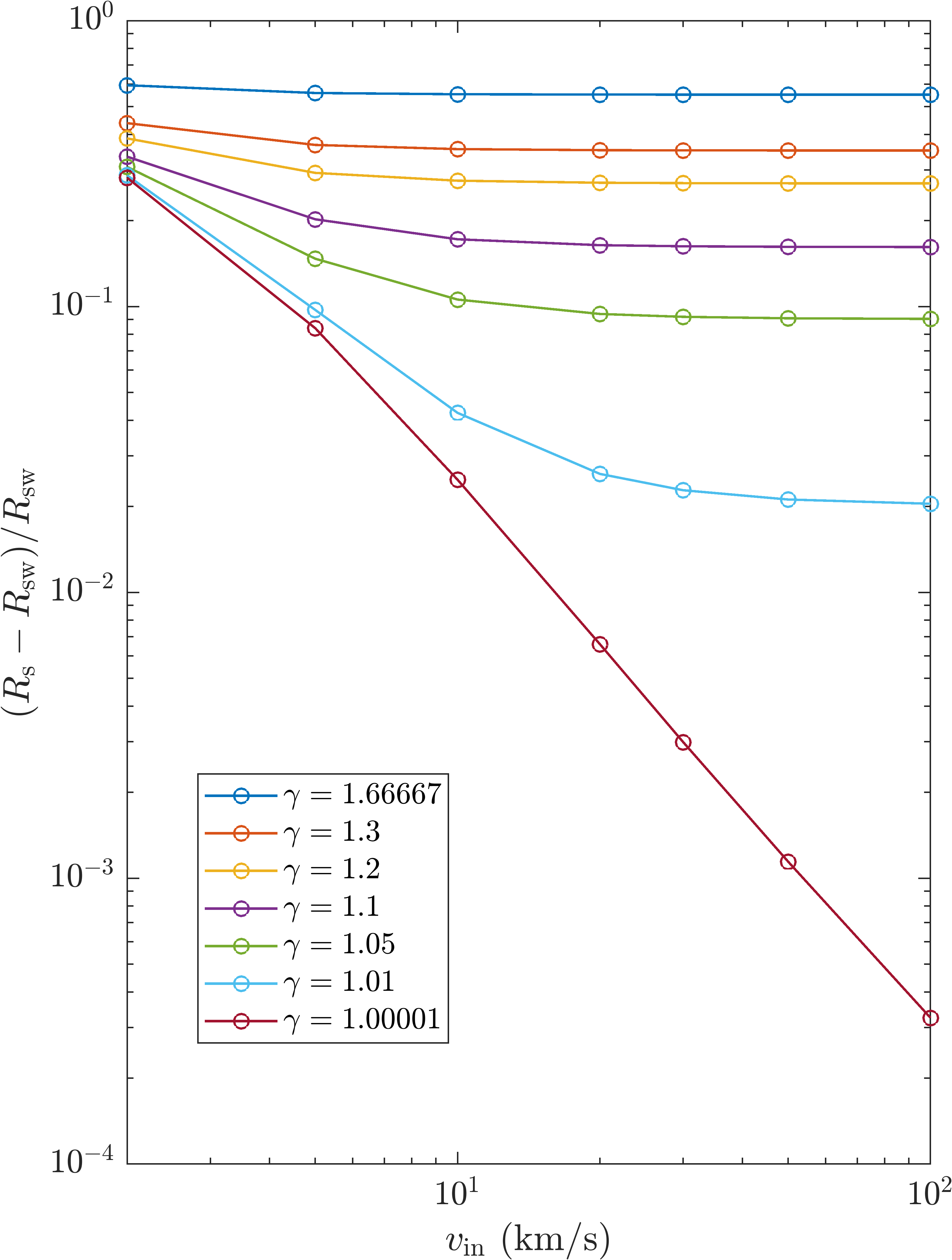}
    \includegraphics[height=2.4in]{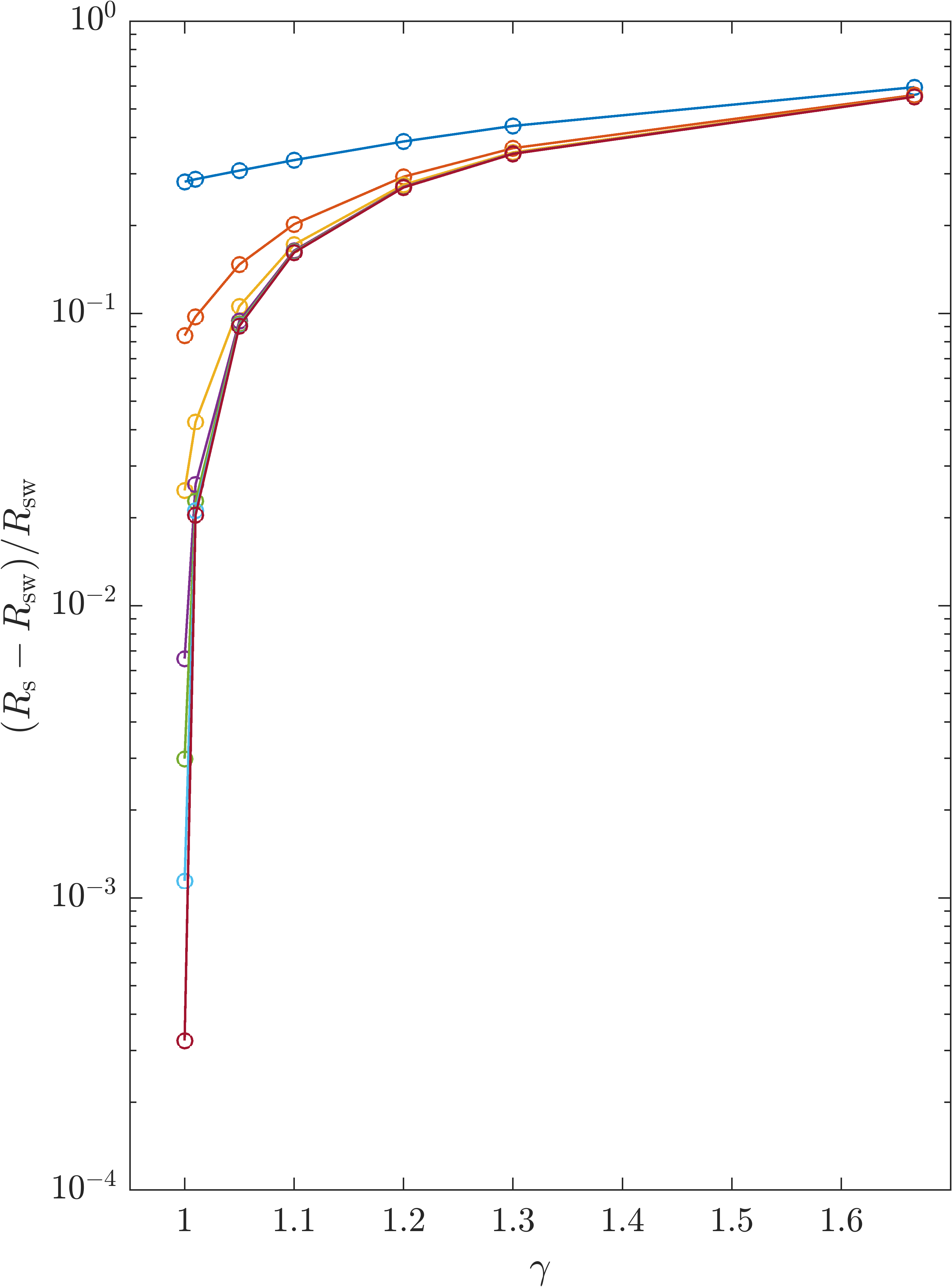}
    \caption{Fractional thickness $(\Rs-\Rsw)/\Rsw$ for the same set as in Figure \ref{fig:thickness}.}
      \label{fig:thicknessm1}
\end{figure}

Figure \ref{fig:thickness} shows the thickness ratios, measured as the relative positions of the discontinuities, $\Rs/\Rsw$ and $\Rc/\Rsw$. The same values are given in Table \ref{tab:thickness}. The related fractional thickness $(\Rs-\Rsw)/\Rsw$ is shown in Figure \ref{fig:thicknessm1}. These quantities are shown as functions of $\gamma$ and $\vin$.
We observe that the efficient dependence of these measures of thickness on $\vin$ decreases with increasing $\gamma$.
Only for the case near the isothermal limit $\gamma=1.00001$ that dependence spans the given $\vin$ values. For $\gamma=1.01$, the effective range of dependence is already limited to $\vin\lesssim10\kms$, and for $\gamma=1.05$ and $1.1$, it is just $\vin\lesssim5\kms$. For $\gamma=1.2$ and $1.3$, clear dependence is seen only for $\vin=2\kms$, and for $\gamma=5/3$, no case of clear dependence of thickness on $\vin$ is observed in this set of simulations.

These results agree with the methods and results of Section \ref{sec:method} for the steeper polytropic EOS and the strongly supersonic values of $\vin$. They also indicate a need for a different treatment for the smallest polytropic $\gamma$ values, using the whole Equation \eqref{eq:compression} instead of its usual strong limit \ref{eq:compression_limit}.
The value of 1.55 found for the thickness ratio $\Rs/\Rsw$ in Table \ref{tab:shell_thickness} agrees completely with the value found in Table \ref{tab:thickness} for $\gamma=5/3$ and large $\vin$ values.

\begin{figure}[ht]
    \centering     
    \includegraphics[height=2.4in]{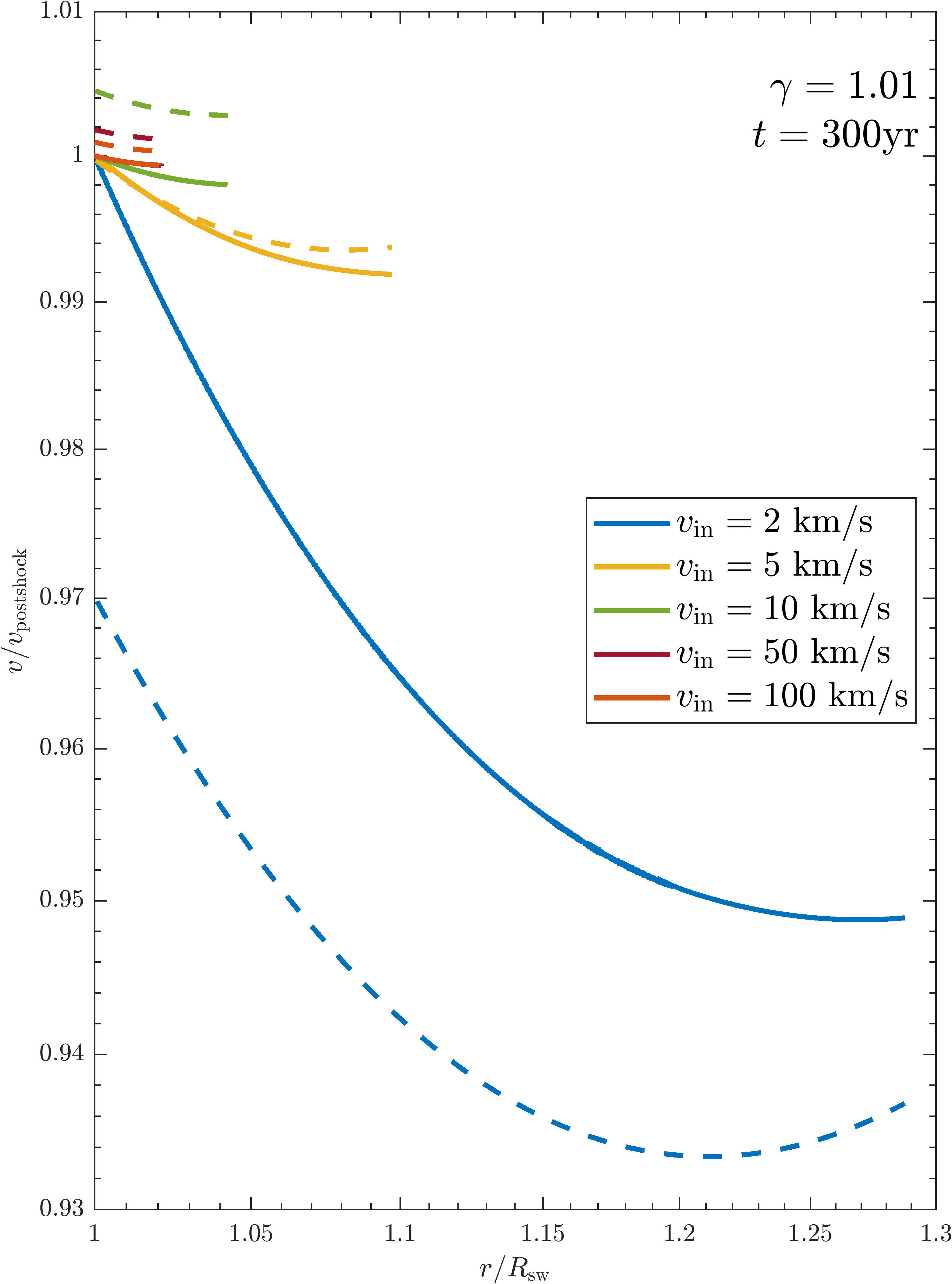}
    \caption{Gas velocity $v$ scaled to the postshock velocity as a function of $r/\Rsw$ at $t=300\yr$ for the small values of $\gamma=1.01$. Solid lines: simulation results. Dashed lines: fits to Equation \eqref{eq:B10}.}
      \label{fig:vel3}
\end{figure}

We also compare our results with one formula also used in Appendix B.2 in KM92b.
Figures \ref{fig:vel1} and \ref{fig:vel2} compare, strictly within the range $\Rsw<r<\Rs$, the profile of $v$ found in the simulations with Equation B10 in KM92b. This equation is derived from mass conservation for an adiabatic pressure. For our case with $\eta=\etain=1$ it is simplified to
\begin{equation}
    \label{eq:B10}
    v(r,t)=\frac{2}{3\gamma}\frac{r}{t}+(1-\frac{2}{3\gamma})\frac{\Rc^3}{r^2 t}\ .
\end{equation}
For this set of simulations, Equation (\ref{eq:B10}) is seen to approximately fit not only the shocked wind ($\Rsw<r<\Rc$) but also the shocked ambient region ($\Rc<r<\Rs$), as seen by the location of the CD in these figures. Comparing the three panels of Figure \ref{fig:vel1}, we see that the lower line ($\vin=2\kms$) shows that the quality of the fit of the numerical results to Equation (\ref{eq:B10}) degrades from $\gamma=1.1$ to $\gamma=1.01$. However, as seen in the log scale of Figure \ref{fig:vel1}, the degraded fit is not entirely meaningless even for the cases in which $\gamma-1\ll 2/M^2$. In Figure \ref{fig:vel3} we rescale the velocity to its numerically-obtained postshock value for $\gamma=1.01$, showing the detail of the degraded fit for $\vin=2\kms$.
The curve shapes resemble the hyperbolas used in Section \ref{sec:matching}.


\begin{table}[ht]
\centering
\begin{tabular}{|r|lllllll|lllllll|}
\hline\hline
\multicolumn{1}{|c|}{ $\vin$ } &
\multicolumn{7}{c|}{$\Rs/\Rsw$ for $\gamma=5/3$ to $1.00001$} &
\multicolumn{7}{c|}{$\Rc/\Rsw$ for $\gamma=5/3$ to $1.00001$} \\ \hline
 (km\,s)
& 5/3 & 1.3 & 1.2 & 1.1 & 1.05 & 1.01 & {\tiny 1.00001}
& 5/3 & 1.3 & 1.2 & 1.1 & 1.05 & 1.01 & {\tiny 1.00001}
 \\ \hline
100 & 1.551 & 1.351 & 1.270 & 1.161 & 1.090 & 1.020 & 1.000 & 1.306 & 1.222 & 1.179 & 1.114 & 1.067 & 1.016 & 1.000 \\
50 & 1.551 & 1.351 & 1.270 & 1.162 & 1.091 & 1.021 & 1.001 & 1.306 & 1.222 & 1.179 & 1.115 & 1.067 & 1.016 & 1.001 \\
30 & 1.551 & 1.352 & 1.270 & 1.162 & 1.092 & 1.023 & 1.003 & 1.306 & 1.222 & 1.179 & 1.115 & 1.068 & 1.017 & 1.003 \\
20 & 1.551 & 1.352 & 1.271 & 1.164 & 1.094 & 1.026 & 1.007 & 1.306 & 1.222 & 1.180 & 1.116 & 1.069 & 1.020 & 1.006 \\
10 & 1.553 & 1.355 & 1.275 & 1.172 & 1.106 & 1.042 & 1.025 & 1.306 & 1.223 & 1.182 & 1.120 & 1.077 & 1.032 & 1.018 \\
5 & 1.558 & 1.367 & 1.293 & 1.202 & 1.147 & 1.097 & 1.084 & 1.306 & 1.227 & 1.189 & 1.137 & 1.103 & 1.071 & 1.061 \\
2 & 1.594 & 1.438 & 1.387 & 1.334 & 1.308 & 1.288 & 1.282 & 1.310 & 1.250 & 1.230 & 1.212 & 1.204 & 1.199 & 1.198 \\ \hline
\end{tabular}
\caption{Thickness ratios. Same data as in Figure \ref{fig:thickness}}
\label{tab:thickness}
\end{table}

\section{Discussion} \label{sec:Discussion}
In this work, we have built a series of results to study wind interactions with an environment. In Section \ref{sec:method}, we designed and applied to study 1D bubble properties using a semi-analytical method.
In Section \ref{sec:vin}, we extended the simulation methods of S20A (\citet{shang2020} App.\ A) to cover a wide range of $\gamma$ and $\vin$.
We now compare the semi-analytical and the numerical results, followed by concrete astrophysical applications.
Our solutions are for now limited to the cases in which $\gammasw=\gammasa$. This has applications to cases with negligible radiative cooling, which allows a single $\gamma$ value in a range including $5/3\geq\gamma\gtrsim1.4$ depending on the gas composition. At the other end, cases of very intense cooling on both wind and ambient sides could be modeled as a small $\gamma\sim1$ representing a locally or globally nearly isothermal condition. Intermediate values of $\gamma$ could be applied to a perhaps hypothetical case in which cooling is equally efficient on wind and ambient sides.

\subsection{Comparison of semi-analytical to numerical results}\label{sec:vin_compare}
An assumption of the present semi-analytical method (\S\ref{sec:method}) is that a strong compression regime applies to both shocks. The numerical method (\S\ref{sec:vin}) does not make this assumption. Its results show that the strong compression assumption is well justified for most of the parameter range of $\gamma$ and $\vin$ values, excluding only, as expected, the cases near the isothermal limit and the cases for very small $\vin$ and small $\gamma$.

Careful comparison between the results of \S\ref{sec:vin}, \S\ref{sec:method}, and S20A clarify the dependence of relative thickness $\Rs/\Rsw$ on $\vin$. Section \ref{sec:matched_solution} found values of $\Rs/\Rsw$ which do not depend on $\vin$. This seemed to contrast with the strong dependence of thickness on $\vin$ found in S20A\@. Section \ref{sec:vin} explains this apparent contrast, by showing that there is a transition between both kinds of results in a way that correlates with the formula $\lfrac{(\gamma+1)}{(\gamma-1+2/M^2)}$ for the compression ratio in the two opposite limits. The strong compression limit (1) $\gamma-1\gg 2/M^2$ is obtained for moderate and large $\gamma$, and leads to no dependence of thickness ratio on $\vin$, as in Section \ref{sec:matched_solution}. The opposite limit, (2) $\gamma-1\ll 2/M^2$, leads to a strong dependence of thickness on $\vin$, a condition that requires a sufficiently small $\gamma$ value, in practice close to the isothermal limit of very intense cooling, as in S20A and in the model with $\gamma=1.00001$ in Section \ref{sec:vin} here. The cases with $\gamma$ between $\sim1.01$ and $1.1$ provide intermediate transitions.
The cases with $\gamma-1\ll 2/M^2$ have application to fully radiative bubbles. Their compression ratios effectively equal the isothermal limit values square $M_\text{preshock}^2$ of their preshock, comoving Mach numbers at each of the two shocks.
The thickness of these bubbles is seen in Figures \ref{fig:vindep1}--\ref{fig:vindep3} to grow with $\vin$. One important concrete example is that of nearly isothermal models of slow winds: the thickness of such winds, measured as the ratio $(\Rs-\Rsw)/\Rs$ can become sensitively dependent on $\vin$.

These cases with very small $\gamma$ are not globally isothermal, because the sound speeds are kept distinct between the free ambient and wind media. A globally isothermal case is presented in Appendix \ref{sec:isothermal}, leading to the different result that the CD disappears.

\begin{figure}[ht]
    \centering     
    \includegraphics[height=2.4in]{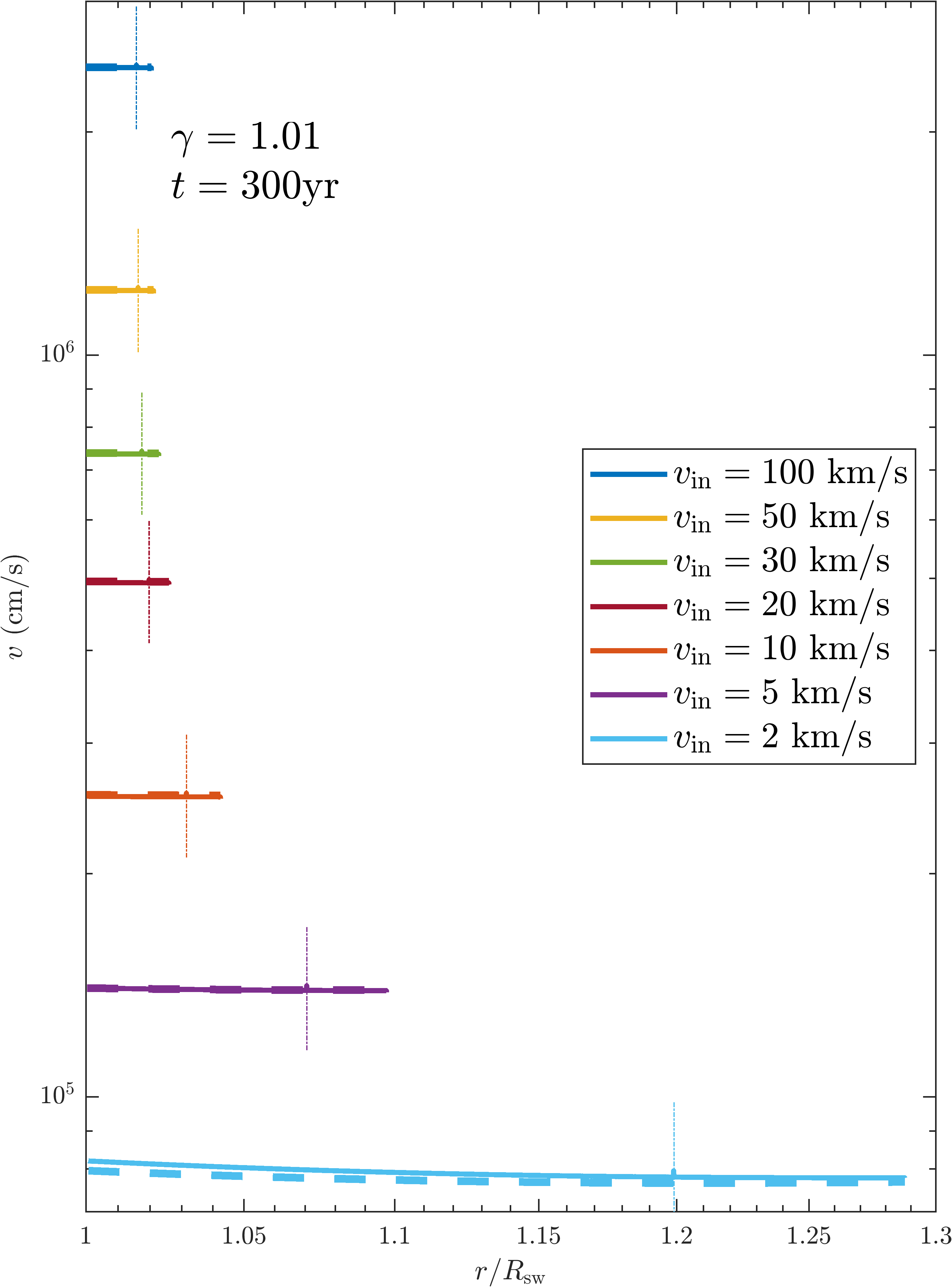}
    \includegraphics[height=2.4in]{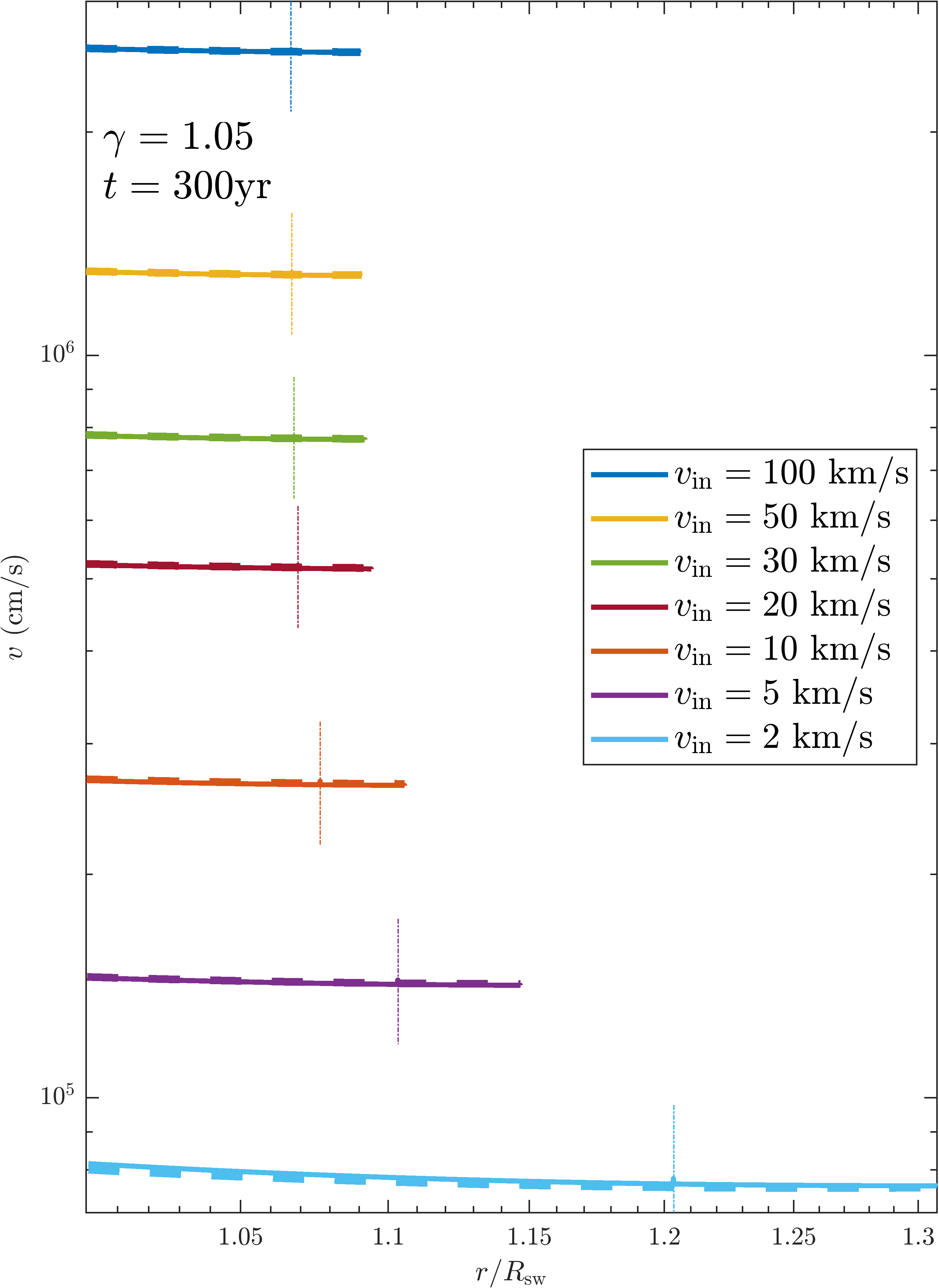}
    \includegraphics[height=2.4in]{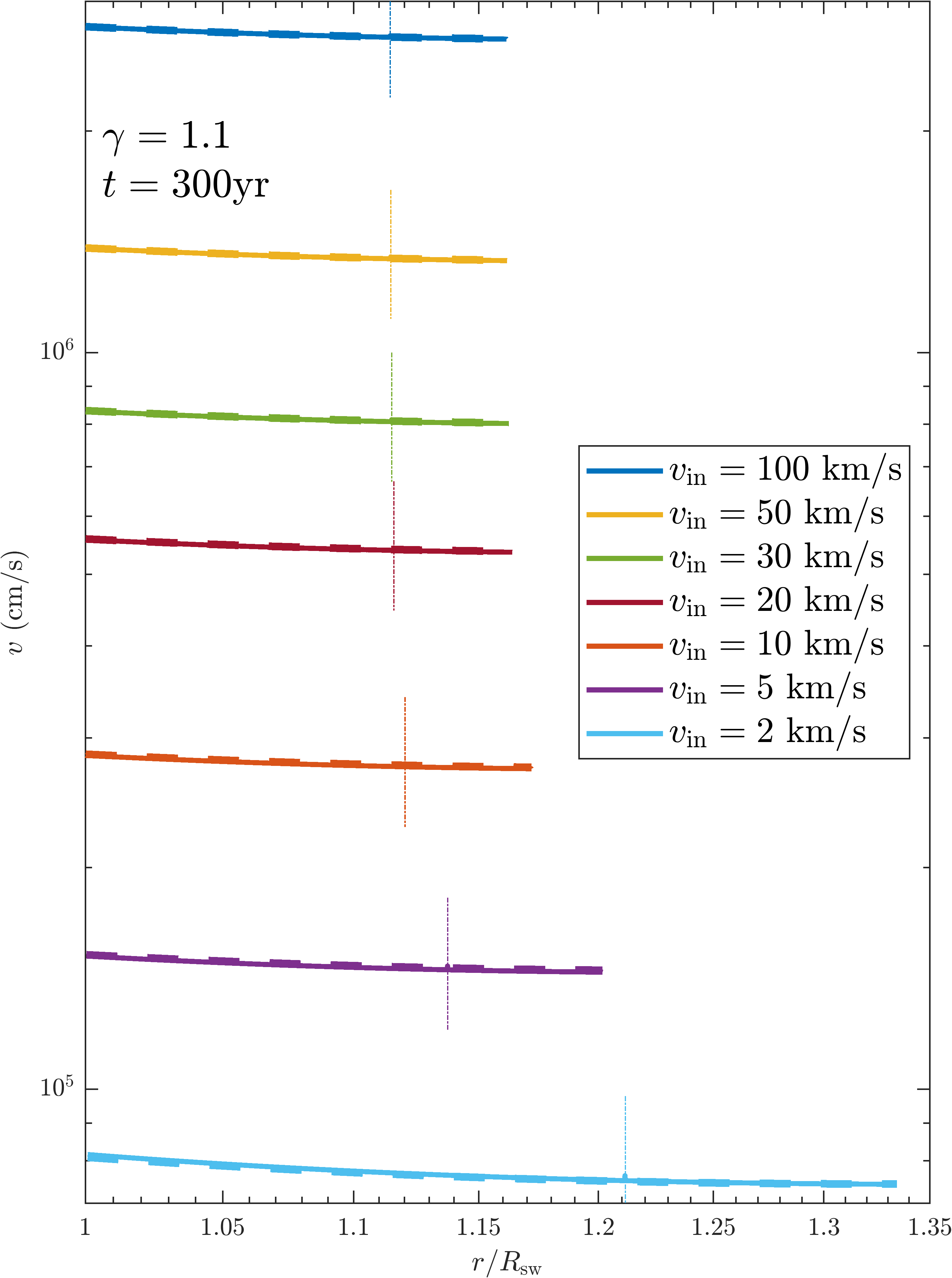}
    \caption{Gas velocity $v$ as a function of $r/\Rsw$ at $t=300\yr$ for small values of $\gamma$. Solid lines: simulation results. Dashed lines: fits to Equation (\ref{eq:B10}). Thin vertical lines: position of the contact discontinuity.}
      \label{fig:vel1}
\end{figure}
\begin{figure}[ht]
    \centering     
    \includegraphics[height=2.4in]{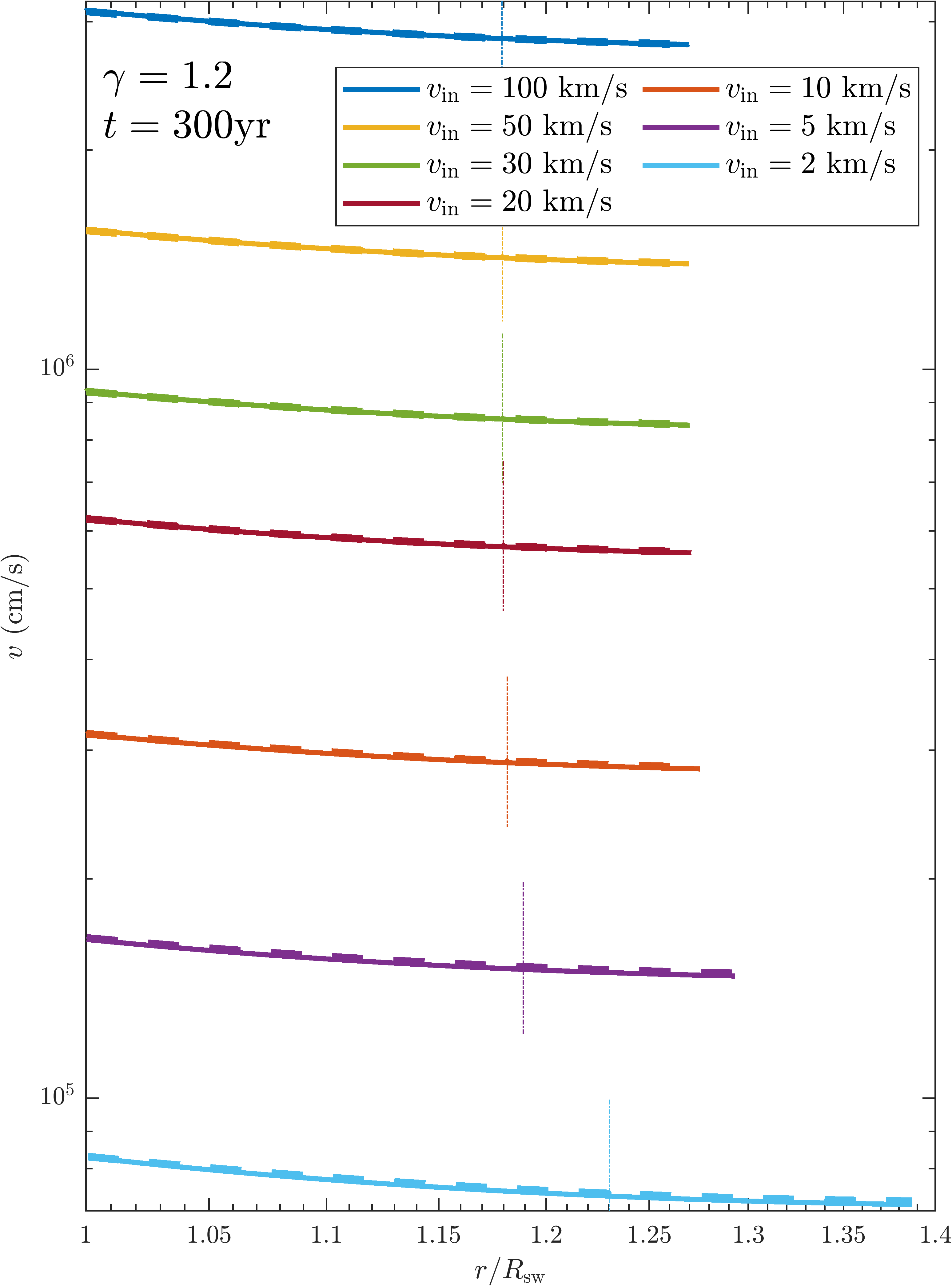}
    \includegraphics[height=2.4in]{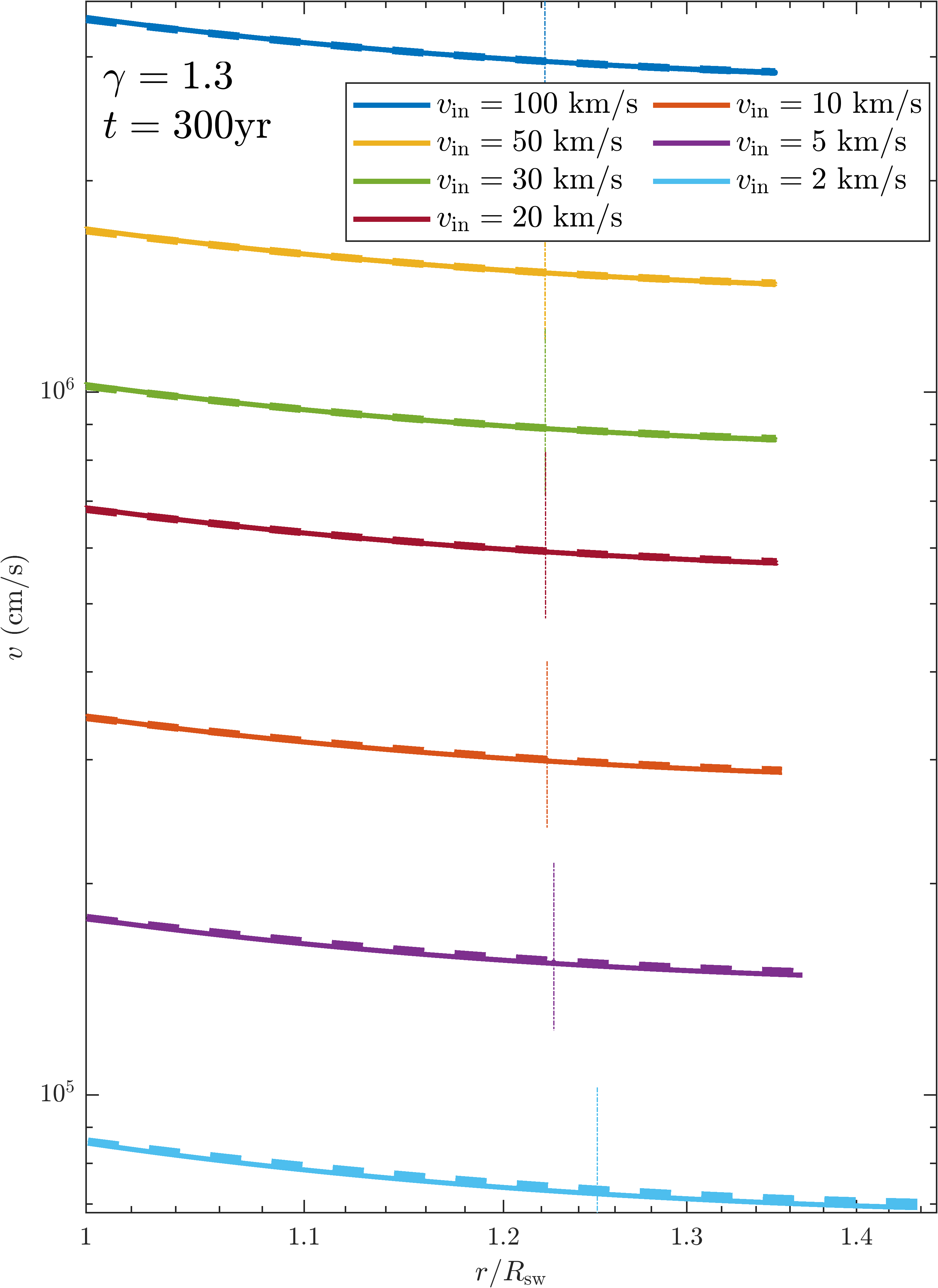}
    \includegraphics[height=2.4in]{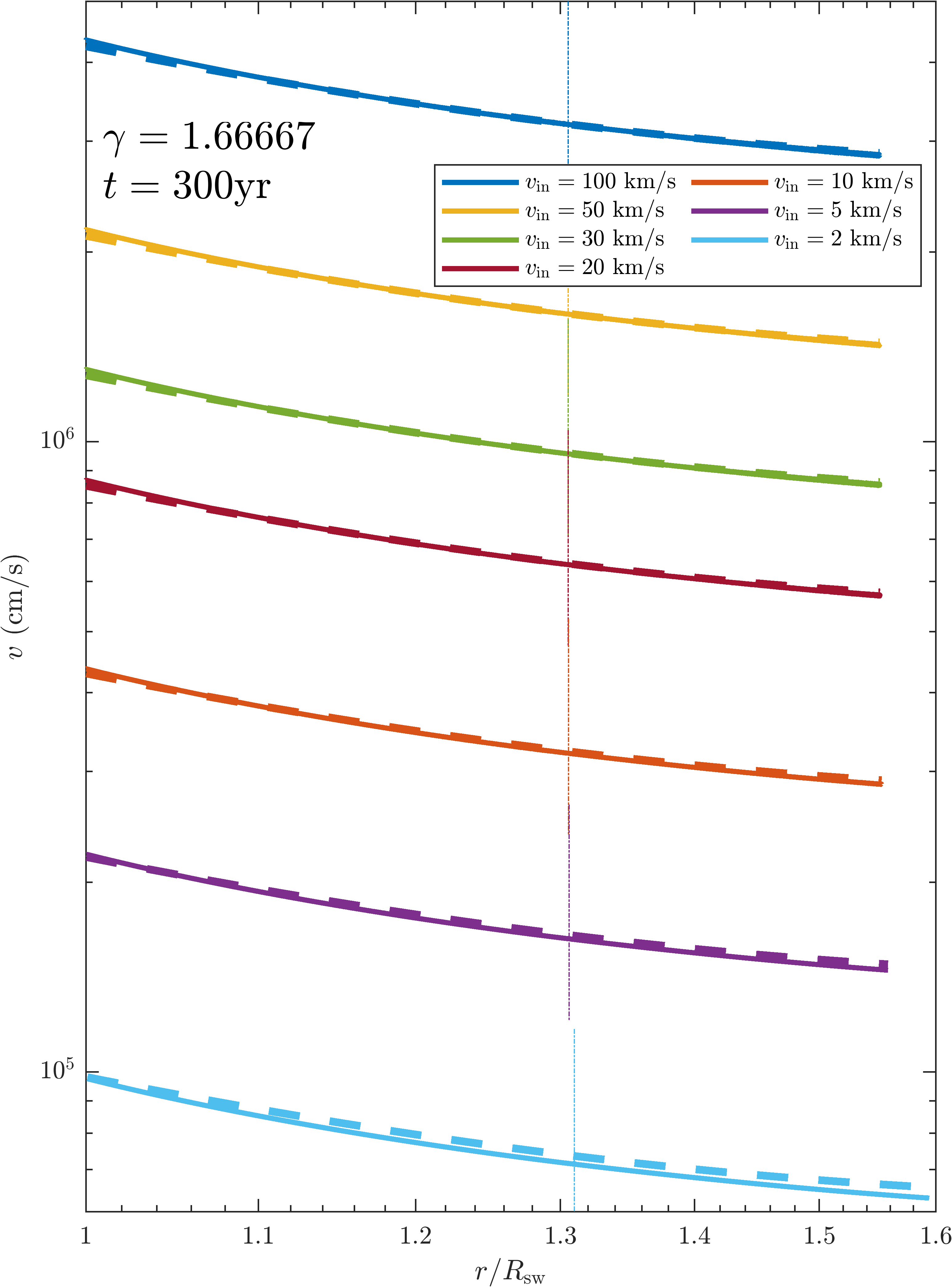}
    \caption{Same as Figure \ref{fig:vel1} for moderate and large values of polytropic $\gamma$.}
      \label{fig:vel2}
\end{figure}

\subsection{Astrophysical applications: Wind flowing into a preexistent wind}\label{sec:applications_wind}
We consider now cases of astrophysical application of the model of this work, with $\etain=1$ and $k_\rho=2$.
Ambient media with $\rho\propto r^{-2}$ are present in various situations.
In this section we over one natural cases with spherical symmetry: a steady wind flowing into an ambient medium formed by the bubble of a preexisting wind, such as in PPNs. In the following section we cover and winds flowing into a young molecular cloud collapse region, in the absence of magnetic field.

We want to utilize the analytical and numerical results of Section \ref{sec:all_results} to the case of a wind flowing into a preexistent wind, whose quantities are described with a subscript $p$.
Equation \eqref{eq:rhow} shows that the free wind region of the preexistent wind has a density profile ${\rhow}_p=C_pr^{-2}\propto r^{-2}$, valid in the region $r<{\Rsw}_p$, where the constant $C_p$ carries information about the mass loading $\Mdotin{}_p$ of the preexistent wind.
This power law ($r^{-2}$) is consistent with the conditions of this work.

However, there are additional constraints. The velocity within that preexisting free wind has the simple velocity profile $v={\vin}_p$, which provides an ambient region to the new wind. 
One constraint is that the model of this work without any generalization requires the free ambient region to have negligible velocity. This can be expressed as ${\vin}_p\ll \vs$, where the ambient shock velocity $\vs$ is a characteristic velocity of the shocked ambient region of the new wind.
We note that $\vs$ is smaller than $\vin$.
This gives the constraint ${\vin}_p\ll \vs < \vin$, requiring that the upcoming wind must be much faster than the preexistent wind.

One additional constraint is that the mass swept up by the outer shock of the upcoming wind must be less than the mass ejected in the preexisting, slow wind. We will show this constraint on masses is equivalent to a constraint on wind radii,
The mass swept up by the outer shock of the upcoming wind is the integral of the density profile of the preexistent wind between $r=0$ and $r=\Rs$. This is equal to the integral of the form
\begin{equation}\label{eq:mass_swept}
    \int^R 4\pi r^2 C_p r^{-2}\,dr=\int^R 4\pi C_p\,dr=4\pi C_pR\ ,
\end{equation}
where $R=\Rs$. The mass ejected in the preexistent wind follows the same Equation \eqref{eq:mass_swept} but now with $R={\Rsw}_p$, where ${\Rsw}_p$ sets the limit of the free wind region of the preexistent wind.
Comparing these two mass integrals then gives the  $4\pi C_p\Rs < 4\pi C_p {\Rsw}_p$, equivalent to the very simple and intuitive constraint $\Rs < {\Rsw}_p$ that the ambient shock of the upcoming wind must be located completely inside the free wind region of the preexistent wind. 

Because the upcoming wind is faster than the preexisting wind, this sets up a natural time constraint: the upcoming shock must be far from having caught up with the wind shock of the preexisting bubble.
This time constraint can be estimated from $t<{\Rsw}_p/(\vin-{\vin}_p)={\Rsw}_p/\vin=t_p{\vsw}_p/\vin$, where $t_p$ is the time since the emergence of the preexisting wind. A natural time scale for this is the time difference $t_p-t$ between preexisting and upcoming wind, multiplied by the factor ${\vsw}_p/\vin$, which must be $\ll1$ because of previously deduced constraints.
PPNs provide examples of bubbles blown into an ambient created by a preexisting wind
\citep[e.g.,][and references therein]{garcia-segura2022}.

\subsection{Astrophysical Applications: Bubbles in young collapsing cores and molecular clouds}\label{sec:applications_YSO}
Collapsing cores can naturally give rise to spherical media with $k_\rho=2$ density power laws.
Under the assumption of spherical symmetry, a theoretical basis for processes producing this power law is presented in 
\citet{bodenheimer1968} and \citet{shu1991}. This theory fits observations of various collapsing cores. For example, the Class 0 protostellar source B335 has been measured to have an $1/r^2$ power law in its large scale \citep[e.g.,][]{zhou1993,saito1999,kurono2013}.
More recently, \citet{lombardi2015} have studied eight molecular clouds with spherical models, and found density distributions with power-law shapes, with a typical exponent within $2\pm0.3$ for their models of Oph, Orion A, Orion B, Perseus, and Taurus (their models for Polaris, California, and Pipe are however steeper). Power laws near 2 are therefore attested to be frequent, but they are however not universal in this sample. They are also subject to the assumption of spherical symmetry, as for Orion A, a model taking into account the filamentary nature of the clouds \citep{stutz2016} found an exponent of $13/8=1.625$.
Steady and fast winds flowing through these ambient media provide a natural application of the $k_\rho=2$ models of this work.

\subsection{Estimations of velocity ratios}\label{sec:vel_ratios}
Our setting $\etain=1$ and $k_\rho=2$ implies that shock velocities $\vs$ and $\vsw$ are constant, and respectively equal to $\Rs/t$ and $\Rsw/t$.
The ratios of these velocities to $\vin$ can then be estimated using results and methods of KM92b as follows.
For our choices of parameters, and using their definition of $\cal L$,  Equation 3.1 of KM92b simplifies to
$\Rs=[(2\pi/3)\Gamma_\text{rad}\xi\delta\vin^3]^{1/3} t$, where the product $\Gamma_\text{rad}\xi$ is as utilized in KM92b, and $\delta$ is the ratio we utilized in Section \ref{sec:matching}. This can be further simplified to the velocity ratio $\vs/\vin=[(2\pi/3)\Gamma_\text{rad}\xi\delta]^{1/3}$. The value of the product $\Gamma_\text{rad}\xi$ has been estimated in Equation 3.10 in KM92b, which for our parameters simplifies to the expression $\Gamma_\text{rad}\xi\approx(9/8\pi)(\gammasw-1)(\gammasa+1)/[(3\gammasw-2)\lambda_c^3]$.
Similarly, Equations 3.14 in KM92b simplifies for our settings to $\Rsw/t=\vsw=(3/f_P)^{1/2}\vin$.
The velocity ratio is then $\vsw/\vin=(3/f_P)^{1/2}$, where the value $f_P$ can be estimated from Equation 3.16 of KM92b as a function of the constants $\gammasw$ and $\gammasa$, which in our setting simplifies to
\begin{equation}
    f_P=\frac{\gammasw+1}{\gammasa+1}\left[\frac{4\gammasw}{(\gammasw+1)^2}\right]^\frac{\gammasw}{\gammasw-1}\ .
\end{equation}

\section{Summary} \label{sec:conclusions}
We have analyzed the interactions between stellar wind and ambient media with $\rho = r^{-2}$, forming wind-driven bubbles. The analytical results in spherically symmetric configurations apply the methods of KM92b, extending its ODE formulation given the Mach number from the shocked ambient material to the treatment of the shocked wind region (Section \ref{sec:method}). We obtained a system of ODEs, based on the physical conservation laws, that can be solved. We have also developed an analytical hyperbolic approximation to the ODEs that is sufficient to be used separately from the hydrodynamic equations. This extended formulation leads to an intuitive estimation of the bubble thickness ratio $\Rs/\Rsw$, a quantity crucial to bubble dynamics and observation, and bubble structure. The results completely agree with numerical methods (Section \ref{sec:vin}) for the usual shock cases in which the Rankine-Hugoniot compression ratios are strong. Section \ref{sec:vin} also studies the dependence on wind velocity, exploring even the regimes of less strong compression ratio with a numerical method, resulting again in strong quantitative agreement for most cases, with the notable exception of very small polytropic $\gamma$ values, especially in the presence of relatively modest wind velocities. While the parameter range of that exception is not large, it includes the important isothermal limit, and it will be the topic of further investigation. In Section \ref{sec:vin_compare} we compare the analytical to the numerical results. Astrophysical applications to various systems are discussed in Sections \ref{sec:applications_wind} and \ref{sec:applications_YSO}. For the considered case $k_\rho = 2$, $\etain = 1$, it is possible to estimate $\vs / \vin$ and $\vin / \vsw$ using methods of KM92, which are discussed in Section \ref{sec:vel_ratios}. A globally isothermal solution for the large Mach number case is presented in Appendix \ref{sec:isothermal}. An analytical solution for an analog Cartesian problem is presented in Appendix \ref{sec:riemann}, showing qualitative agreement of shock structure.

The scope of this work is the proof of concept of the method to study bubbles within its assumptions, which we have shown to be valid within the parameter range explored. Future works along this line would explore a wider parameter space for bubbles belonging to different regimes.

Our method widely applies to spherically symmetric systems based on the above assumptions. It does not explicitly consider the magnetic effects; however, it can be used as a hydrodynamic baseline for a wide range of bubble spherically symmetric objects connected to supernova remnants and evolved planetary nebulae. The relative thickness of the bubble shells (Table \ref{tab:shell_thickness}, Table \ref{tab:shell_thickness_mach} and Figure \ref{fig:thickness}) is very useful information to improve our understanding of astrophysical processes of observed sources by connecting the inner scale of $\Rsw$ to the outer scale of $\Rs$, which may have different conditions for real observations.

\begin{acknowledgments}
The authors acknowledge support for the CompAS Project from the Institute of Astronomy and Astrophysics, Academia Sinica (ASIAA),  the Academia Sinica grant AS-IAIA-114-M01, and the National Science and Technology Council (NSTC) in Taiwan through grants 112-2112-M-001-030, 113-2112-M-001-008, and 113-2927-I-001-513-. DZ acknowledges support for an internship at the ASIAA through the International Internship Pilot Program (IIPP) funded by NSTC\@. The authors thank the National Center for High-performance Computing (NCHC) of National Applied Research Laboratories (NARLabs) in Taiwan for providing computational and storage resources and the ASIAA for in-house access to high-performance computing facilities.

\end{acknowledgments}

\appendix
\section{Isothermal bubble with large Mach number}\label{sec:isothermal}
An exactly isothermal bubble has a simplified bubble structure without a contact discontinuity. The globally constant value of the sound speed makes the transition of both pressure and density smooth between the compressed wind and the compressed ambient medium. The compression ratio at each of the two shocks is equal to their respecive comoving $M_\text{preshock}^2$ value. At the wind shock, the comoving Mach numbers are $M_\text{preshock}=|\vin-\vsw|/a$, and at the ambient shock it is $M_\text{preshock}=|0-\vs|/a$, where $a\equiv\aamb=\awind$.
We use (as in \citealt{shu1977} and \citealt{deschner2018}) a dimensionless radial coordinate $x\equiv r / a t$ (with $\xsw=\Rsw/a t$ and $\xs=\Rs/at$) and a dimensionless velocity $u\equiv v/a$.
The dimensionless preshock velocities are then respectively $\vin/a$ and $0$, and their corresponding postshock velocities are then
\begin{align}
    u_\text{postshock,w}&=\xsw+\frac{1}{\uin-\xsw},\\
    u_\text{postshock,a}&=\xs-\frac{1}{\xs}\ .
\end{align}
We assume that the ambient density is $\propto r^{-2}$, and then the ODE for the dimensionless $u$ can be written as
\begin{equation}
    \frac{du}{dx} = \frac{2(u-x)}{x[(x-u)^2-1]}
\end{equation}
an ODE consistent with both \citet{shu1977} and \citet{deschner2018}, although our boundary conditions are substantially different. In this appendix we present an analytical solution valid for the case in which the Mach number $\vin/a$ of the isothermal bubble is very large.
Under these conditions, the equations above simplify considerably, because now the shell becomes thin, and
$\uin\gg \xs \gg 1 \gg (\xs-\xsw)\equiv \Delta\xs$, with $\Delta\ll1$ and $(x-u)^2\ll 1$ in the thin shell region between shocks.
The approximate location of the thin shell can be found in the cold limit from the usual momentum-conserving formula $\vs=\vin/(1+\delta^{-1/2})$ (KM92b Equations A3, A5).
The difference between the two postshock velocities is very small, of the order of 
\begin{equation}
    \frac{du}{dx} (\xs-\xsw) = \frac{2(u-x)}{x} (\xs-\xsw)\ ,
\end{equation}
which is $\ll$ than either of the two postshock velocities.
We then set up the approximate equation that the two postshock velocites are equal, and deduce a value for the relative shock thickness $\Delta=(\Rs-\Rsw)/\Rs$,
\begin{align}
    &\xs-\frac{1}{\xs}=\xsw+\frac{1}{\uin-\xsw}\\
    &\Delta\xs-\frac{1}{\xs}=\frac{1}{\uin-\xs-\Delta\xs}\approx\frac{1}{\uin-\xs}\\
    &\Delta\approx\frac{1}{\xs} \left(\frac{1}{\uin - \xs} + \frac{1}{\xs}\right)\label{eq:Delta}\ .
\end{align}
This estimate depends on the sound speed $a$, but for a case in which the Mach number $\vin/a$ is very large.
We have computed numerically a globally isothermal case with $\vin=30\kms$ and $a=0.2\kms$, which has a large Mach number. The results, in sufficient agreement, are shown in Table \ref{tab:shell_thickness2}.
These models can be applied to certain kinds of intensely radiative bubbles, extended enough that a single temperature applies to both wind and ambient regions, perhaps involving turbulence within the intershock region. A generalization to two-temperature radiative bubbles seems probably feasible, adding however one free parameter to the equations, which may increase the complexity of the dimensional arguments applied here. That free parameter, the temperature ratio, will be solved at the contact discontinuity of the generalized setting.
\begin{table}[ht]
\centering
\begin{tabular}{*{2}c}
\hline\hline
Numerical $\Delta$ & Eq.\ \eqref{eq:Delta} \\ \hline
0.12\% & 0.10\% \\ \hline
\end{tabular}
\caption{Comparison of relative shell thickness calculated from a numerical result and using Eq.\ (\ref{eq:Delta}).}
\label{tab:shell_thickness2}
\end{table}

\section{Riemann Problem for Euler Equations} \label{sec:riemann}
\begin{figure}[ht]
    \centering
    \includegraphics[width=\linewidth]{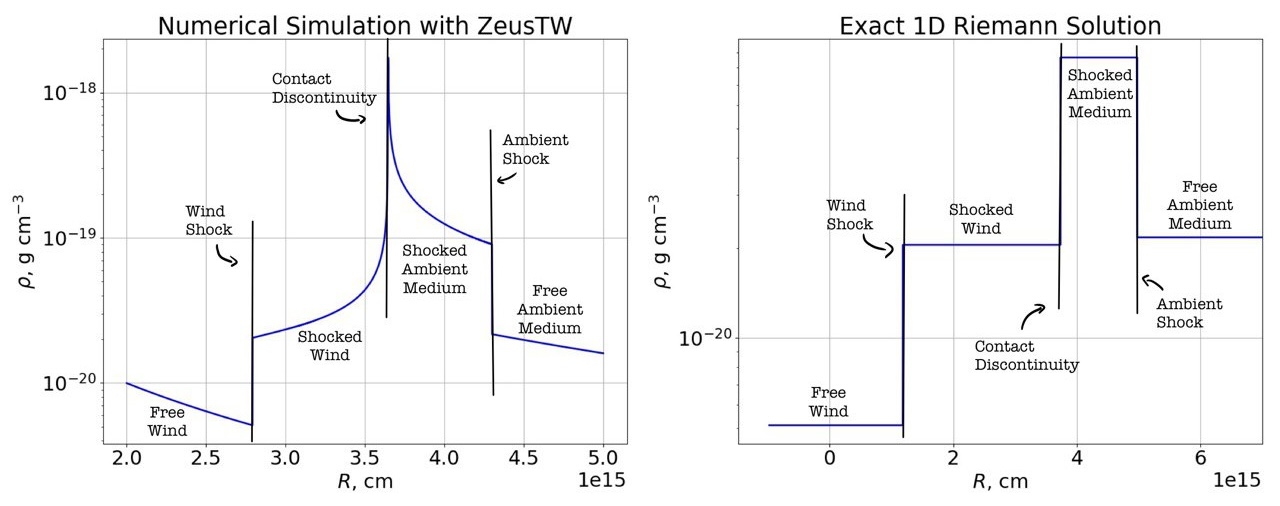}
    \caption{Structures of spherical (left figure) and Cartesian (right figure) bubbles shown on the density profile. The two main bubble regions are shown (free and shocked wind and ambient media), separated by the three classical discontinuities (wind and ambient shock, and a contact discontinuity). These regions are present in both the spherical bubble (left panel: simulated numerically from the PDEs) and in the exact 1D Cartesian solution (right panel: values found using an exact solution of the Riemann problem).
    }
    \label{fig:bubble_structure}
\end{figure}

In this section, we leave the approach of KM92b, on which the previous section is based, for qualitative comparison of the spherical bubble solutions (e.g.\ obtained by high-resolution 1D numerical simulations using \textsc{ZeusTW}) to a well-known analytical result from hydrodynamics. As we see further, both solutions will share the qualitative structure (Fig.\ \ref{fig:bubble_structure}) and physical properties with minor changes.

The Riemann problems in hydrodynamics can be defined as a one-dimensional initial value problem in which the initial condition involves two different constant states separated by a discontinuity. It is regularly applied to studying nonlinear equations, such as conservation laws in hydrodynamics and MHD.

Here, we apply it to a classic example, a 1D Cartesian shock geometry. The two initially constant states of this Riemann problem, the left and right states, are intended to represent the wind and the ambient medium of our bubbles, respectively. The solutions regularly present shocks and discontinuities: a formal mathematical presentation will describe such solutions as \textit{weak solutions} of the system of equations.
For the Euler equations of hydrodynamics in a 1D Cartesian geometry, the exact solution starting from a combination of left states and right states is known \citep[e.g.][]{toro2013,leveque1998}. This solution is known to be self-similar \citep[][]{landau1987}. A critical application of the Riemann problem is to serve as a building block in many computational fluid dynamics (CFD) methods, as pioneered in \citet{godunov1959}, representing a significant fraction of the CFD algorithms and codes.

More formally, the Riemann problem for the one-dimensional time-dependent hydrodynamic equations, by definition, is the initial value problem for conservation laws (\ref{eq: hydro_i}-\ref{eq: hydro_iii}) with the following initial conditions
\begin{equation}
    \vU(x, 0) =
    \begin{cases}
        \vU_\text{L} & \text{if } x < 0, \\
        \vU_\text{R} & \text{if } x > 0,
    \end{cases}
    \text{ where } \vU = \begin{pmatrix} \rho \\ u \\ E \end{pmatrix}.
\end{equation}

In the particular context of this work, we assume initial values $\vU_\text{L}$ and $\vU_\text{R}$ respectively representing the free wind and the free ambient medium regions, letting the self-similar evolution of the Riemann problem represent their interaction.
We utilized the exact forms of the algorithm and program given in \citet{toro2013}. We tested the program on several cases: the Sod shock-tube problem (left rarefaction and right shock, \citealt{sod1978}), the 123 problems (two rarefactions \citealt{eindeldt1991}), the left and right halves of the blast wave problem (left rarefaction and right shock, right rarefaction and left shock, \citealt{woodward1984}), and one test case most interesting to us, involving two shocks. The tests of the program succeeded, and the obtained results were identical to the ones presented in \citet{toro2013}.

After these tests, we used the exact 1D Cartesian solution for comparison with the spherical \textsc{ZeusTW} simulation in high resolution (300,000 radial cells in 1D). Qualitatively, results for the same set of parameters match, as illustrated in Figure \ref{fig:bubble_structure}. Both solutions share the following structure (from left to right): free wind, wind shock (reverse shock), compressed wind, contact discontinuity, compressed ambient medium, ambient shock (forward shock), and free ambient medium.

\begin{figure}[ht]
    \centering
    \includegraphics[width=\linewidth]{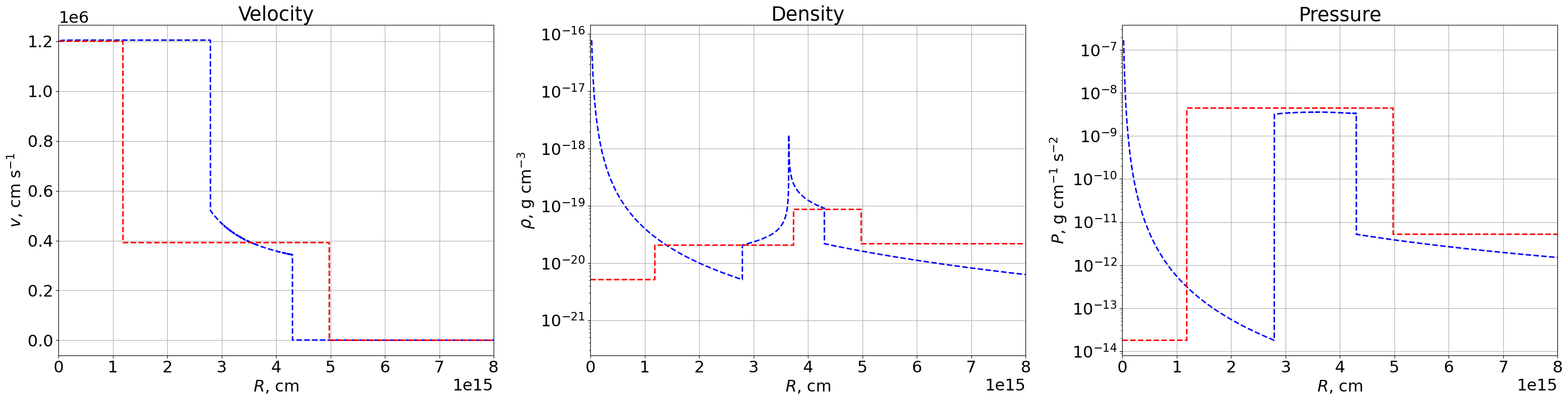}
    \caption{Velocity, pressure and density profiles at 300 years for spherical (blue, \textsc{ZeusTW)} and Cartesian (red, exact 1D solution) bubbles. Spherical solution was obtained as a \textsc{ZeusTW} simulation with the following parameters: $\gamma = 5/3$, $v_\text{L} = 1.2 \times 10^6\cms$, $v_\text{R} = 0\cms$, $a_\text{L} = 6 \times 10^4\cms$, $a_\text{R} = 2 \times 10^6\cms$, $r^2\rho_\text{L} = 4 \times 10^{10}\rsqmassden$, $r^2\rho_\text{R} = 4 \times 10^{11}\rsqmassden$. Cartesian solution was done with the following parameters: $\gamma = 5/3$, $v_\text{L} = 1.2\times 10^6\cms$, $v_\text{R} = 0\cms$, $a_\text{L} = 2412\cms$, $a_\text{R} = 2\times 10^4\cms$, $\rho_\text{L} = 5.12 \times 10^{-21}\massden$, $\rho_\text{R} = 2.18 \times 10^{-20}\massden$.}
    \label{fig:toro_zeus_comparison}
\end{figure}

Figure \ref{fig:toro_zeus_comparison} shows a more detailed comparison. There, we can see that the position of the contact discontinuity approximately matches for both solutions. Still, the positions of the shocks and the thicknesses of the bubbles differ due to obvious differences in geometry. Velocities of free wind and free ambient medium match exactly, velocity in the Cartesian bubble shell stays in range of velocities of spherical bubble shell. Immediate preshock and postshock density values match for both solutions. The qualitative difference of density and pressure in the free wind and free ambient medium regions is also due to different geometry. Interestingly, pressure in the bubble shell is approximately the same in both bubbles. Additionally, pressure is almost constant for the spherical bubble, and isobaric assumption can be used (\citealt{weaver1977}, KM92b).

\bibliographystyle{aasjournal}
\bibliography{main}
\end{document}